\documentclass[twocolumn,tighten,times]{aastex631}

\PassOptionsToPackage{usenames,dvipsnames}{xcolor}
\usepackage{tabularx}
\usepackage{amsmath}  
\usepackage{amssymb}  
\usepackage[T1]{fontenc}
\usepackage{xspace}
\defcitealias{kenworthy_salt3_2021}{K21}

\usepackage{newunicodechar,graphicx}
\DeclareRobustCommand{\okina}{%
  \raisebox{\dimexpr\fontcharht\font`A-\height}{%
    \scalebox{0.8}{`}%
  }%
}
\newunicodechar{ʻ}{\okina}

\newif{\ifchangetext}
\changetextfalse

\InputIfFileExists{\jobname-options}

\ifchangetext
  
  \newcommand{\changenote}[1]{\textcolor{blue}{ \bf #1}}x
\else
  
  \newcommand{\changenote}[1]{}
\fi

\hypersetup{
    colorlinks=True,
    citecolor=black,
    linkcolor=black,
    filecolor=magenta,      
    urlcolor=cyan,
}

\newcommand{\angstrom}{\mbox{\normalfont\AA}\xspace}

\def\arcsec{\ensuremath{^{\prime\prime}}}

\def\GLEE{\textsc{Glee}\xspace}
\def\lenstronomy{\textsc{Lenstronomy}\xspace}
\newcommand{\sersic}{S\'{e}rsic\xspace}

\def\xone{\ensuremath{x_{1}}}
\def\C{\ensuremath{\mathcal{C}}}

\newcommand{\lensedsn}{SN Zwicky\xspace}

\newcommand{\sntd}{{\fontfamily{qcr}\selectfont{SNTD}}\xspace}

\newcommand{\hst}{\textit{HST}\xspace}

\newcommand{\snz}{0.3544}
\newcommand{\lensz}{0.22615}
\def\snstretch{1.16\xspace}
\def\snc{0.005\xspace}
\def\yscode{{\tt lfit\_gui}\xspace}
\def\snpeak{59808.6\xspace}
\begin{document}

\title{LensWatch: I. Resolved \textit{HST} Observations and Constraints on the Strongly-Lensed Type Ia Supernova 2022qmx (``SN Zwicky'')}

\author[0000-0002-2361-7201
]{J.~D.~R.~Pierel}
\correspondingauthor{J.~D.~R.~Pierel} 
\email{jpierel@stsci.edu}
\affil{Space Telescope Science Institute, 3700 San Martin Drive, Baltimore, MD 21218, USA}
\author[0000-0001-5409-6480]{N.~Arendse}
\affil{Oskar Klein Centre, Department of Physics, Stockholm University, SE-106 91 Stockholm, Sweden}
\author[0000-0002-5085-2143]{S.~Ertl}
\affil{Max-Planck-Institut f{\"u}r Astrophysik, Karl-Schwarzschild Stra{\ss}e 1, 85748 Garching, Germany} \affil{Technische Universit{\"a}t M{\"u}nchen, TUM School of Natural Sciences, Physik-Department, James-Franck-Stra{\ss}e 1, 85748 Garching, Germany}

\author[0000-0001-8156-0330]{X.~Huang}
\affil{Department of Physics \& Astronomy, University of San Francisco, San Francisco, CA 94117, USA}
\affil{Physics Division, Lawrence Berkeley National Laboratory, 1 Cyclotron Road, Berkeley, CA 94720, USA}
\author[0000-0003-3030-2360]{L.~A.~Moustakas}
\affil{Jet Propulsion Laboratory, California Institute of Technology, 4800 Oak Grove Dr, Pasadena, CA 91109, USA}
\author[0000-0003-2497-6334]{S.~Schuldt}
\affil{Dipartimento di Fisica, Universit\`a  degli Studi di Milano, via Celoria 16, I-20133 Milano, Italy}
\author[0000-0002-5558-888X]{A.~J.~Shajib}
\affil{Department of Astronomy and Astrophysics, University of Chicago, Chicago, IL 60637, USA}
\affil{Kavli Institute for Cosmological Physics,  University of Chicago, Chicago, IL 60637, USA}
\author{Y.~Shu}
\affil{Purple Mountain Observatory, No. 10 Yuanhua Road, Nanjing 210033, China}
\author[0000-0003-3195-5507]{S.~Birrer}
\affil{Kavli Institute for Particle Astrophysics and Cosmology \& Department of Physics, Stanford University, Stanford, CA 94305, USA}
\affil{SLAC National Accelerator Laboratory, Menlo Park, CA, 94025, USA}
\affil{Department of Physics and Astronomy, Stony Brook University, Stony Brook, NY 11794, USA}
\author[0000-0002-1537-6911]{M.~Bronikowski}
\affil{Centre for Astrophysics and Cosmology, University of Nova Gorica, Vipavska 11c, Ajdov\v{s}\v{c}ina, Slovenia}
\author[0000-0002-4571-2306]{J.~Hjorth}
\affil{DARK, Niels Bohr Institute, University of Copenhagen, Jagtvej 128, 2200 Copenhagen, Denmark}

\author[0000-0001-5568-6052]{S. H.~Suyu}
\affil{Max-Planck-Institut f{\"u}r Astrophysik, Karl-Schwarzschild Stra{\ss}e 1, 85748 Garching, Germany}
\affil{Technische Universit{\"a}t M{\"u}nchen, TUM School of Natural Sciences, Physik-Department, James-Franck-Stra{\ss}e 1, 85748 Garching, Germany}

\affil{Academia Sinica Institute of Astronomy and Astrophysics (ASIAA), 11F of ASMAB, No.1, Section 4, Roosevelt Road, Taipei 10617, Taiwan}
\author[0000-0002-2350-4610]{S.~Agarwal}
\affil{Department of Astronomy, University of California, Berkeley, CA 94720-3411, USA}

\author[0000-0001-9775-0331]{A.~Agnello}
\affil{DARK, Niels Bohr Institute, University of Copenhagen, Jagtvej 128, 2200 Copenhagen, Denmark}

\author[0000-0002-9836-603X]{A.~S.~Bolton}
\affil{NSF's NOIRLab, 950 N Cherry Ave, Tucson, AZ 85719 USA}

\author{S. Chakrabarti}
\affil{School of Physics and Astronomy, University of Alabama, Huntsville, 301 Sparkman Drive, Huntsville, AL 35899, USA}

\author[0000-0001-7666-1874]{C.~Cold}
\affil{DARK, Niels Bohr Institute, University of Copenhagen, Jagtvej 128, 2200 Copenhagen, Denmark}

\author{F.~Courbin}
\affil{Institute of Physics, Laboratory of Astrophysics, Ecole Polytechnique
Fédérale de Lausanne (EPFL), Observatoire de Sauverny, 1290 Versoix, Switzerland}

\author[0000-0003-0928-2000]{J.~M.~Della~Costa}
\affil{San Diego State University, San Diego, California, 92182}

\author{S.~Dhawan}
\affil{Institute of Astronomy and Kavli Institute for Cosmology, University of Cambridge, Madingley Road, Cambridge CB3 0HA, UK}

\author{M.~Engesser}
\affil{Space Telescope Science Institute, 3700 San Martin Drive, Baltimore, MD 21218, USA}

\author[0000-0003-2238-1572]{Ori~D.~Fox}
\affil{Space Telescope Science Institute, 3700 San Martin Drive, Baltimore, MD 21218, USA}

\author[0000-0002-8526-3963]{C.~Gall}
\affil{DARK, Niels Bohr Institute, University of Copenhagen, Jagtvej 128, 2200 Copenhagen, Denmark}

\author[0000-0001-6395-6702]{S.~Gomez}
\affil{Space Telescope Science Institute, 3700 San Martin Drive, Baltimore, MD 21218, USA}

\author[0000-0002-4163-4996]{A.~Goobar}
\affil{Oskar Klein Centre, Department of Physics, Stockholm University, SE-106 91 Stockholm, Sweden}

\author[0000-0001-8738-6011]{S.~W.~Jha}
\affil{Department of Physics \& Astronomy, Rutgers, State University of New Jersey, 136 Frelinghuysen Road, Piscataway, NJ 08854, USA}

\author[0000-0003-3100-7718]{C.~Jimenez}
\affil{Instituto de Astrofis\'{i}ca de Canarias, 38200 La Laguna,Tenerife, Canary Islands, Spain}

\author[0000-0001-5975-290X]{J.~Johansson}
\affil{Oskar Klein Centre, Department of Physics, Stockholm University, SE-106 91 Stockholm, Sweden}

\author{C.~Larison}
\affil{Department of Physics \& Astronomy, Rutgers, State University of New Jersey, 136 Frelinghuysen Road, Piscataway, NJ 08854, USA}

\author{G.~Li}
\affil{Purple Mountain Observatory, No. 10 Yuanhua Road, Nanjing 210033, China}

\author[0000-0001-8442-1846]{R.~Marques-Chaves}
\affil{Department of Astronomy, University of Geneva, 51 Chemin Pegasi, 1290 Versoix, Switzerland}

\author[[0000-0001-8317-2788]{S.~Mao}
\affil{Department of Astronomy, Tsinghua University, Beijing 100084, China}

\author{P.~A.~Mazzali}
\affil{Astrophysics Research Institute, Liverpool John Moores University, IC2, Liverpool Science Park, 146 Brownlow Hill, Liverpool L3 5RF, UK}
\affil{Max-Planck-Institut f{\"u}r Astrophysik, Karl-Schwarzschild Stra{\ss}e 1, 85748 Garching, Germany}

\author[0000-0002-2807-6459]{I.~Perez-Fournon}
\affil{Instituto de Astrofis\'{i}ca de Canarias, 38200 La Laguna,Tenerife, Canary Islands, Spain}
\affil{Departamento de Astrof\'{\i}sica, Universidad de La Laguna (ULL), 38206 La Laguna, Tenerife, Spain}

\author[0000-0003-4743-1679]{T.~Petrushevska}
\affil{Centre for Astrophysics and Cosmology, University of Nova Gorica, Vipavska 11c, Ajdov\v{s}\v{c}ina, Slovenia}

\author[0000-0002-5391-5568]{F.~Poidevin}
\affil{Instituto de Astrofis\'{i}ca de Canarias, 38200 La Laguna,Tenerife, Canary Islands, Spain}
\affil{Departamento de Astrof\'{\i}sica, Universidad de La Laguna (ULL), 38206 La Laguna, Tenerife, Spain}

\author[0000-0002-4410-5387]{A.~Rest}
\affil{Space Telescope Science Institute, 3700 San Martin Drive, Baltimore, MD 21218, USA}
\affil{Physics and Astronomy Department, Johns Hopkins University, Baltimore, MD 21218, USA.}

\author[0000-0003-1889-0227]{W.~Sheu}
\affil{Department of Physics and Astronomy, University of California, Los Angeles, 430 Portola Plaza, Los Angeles, CA 90095, USA}

\author[0000-0002-1114-0135]{R.~Shirley}
\affil{School of Physics and Astronomy, University of Southampton, Southampton, United Kingdom}

\author{E.~Silver}
\affil{Department of Astronomy, University of California, Berkeley, CA 94720-3411, USA}

\author[0000-0002-0385-0014]{C.~Storfer}
\affil{Institute for Astronomy, University of Hawaii, Honolulu, HI 96822}

\author[0000-0002-7756-4440]{L.~G.~Strolger}
\affil{Space Telescope Science Institute, 3700 San Martin Drive, Baltimore, MD 21218, USA}

\author[0000-0002-8460-0390]{T. Treu}
\affiliation{Department of Physics and Astronomy, University of California, Los Angeles, 430 Portola Plaza, Los Angeles, CA 90095, USA}

\author[0000-0001-9666-3164]{R.~Wojtak}
\affil{DARK, Niels Bohr Institute, University of Copenhagen, Jagtvej 128, 2200 Copenhagen, Denmark}

\author[0000-0002-0632-8897]{Y.~Zenati}
\affil{Physics and Astronomy Department, Johns Hopkins University, Baltimore, MD 21218, USA.}

\begin{abstract}
Supernovae (SNe) that have been multiply-imaged by gravitational lensing are rare and powerful probes for cosmology. Each detection is an opportunity to develop the critical tools and methodologies needed as the sample of lensed SNe increases by orders of magnitude with the upcoming Vera C. Rubin Observatory and \textit{Nancy Grace Roman Space Telescope}. The latest such discovery is of the quadruply-imaged Type Ia SN 2022qmx \citep[aka, ``SN Zwicky'';][]{goobar_sn_2022} at $z=\snz$. \lensedsn was discovered by the Zwicky Transient Facility (ZTF) in spatially unresolved data. Here we present follow-up \textit{Hubble Space Telescope} observations of \lensedsn, the first from the multi-cycle ``\href{www.lenswatch.org}{LensWatch}'' program. We measure photometry for each of the four images of \lensedsn, which are resolved in three WFC3/UVIS filters (F475W, F625W, F814W) but unresolved with WFC3/IR~F160W, and present an analysis of the lensing system using a variety of independent lens modeling methods. We find consistency between lens model predicted time delays ($\lesssim1$ day), and delays estimated with the single epoch of \hst colors ($\lesssim3.5$ days), including the uncertainty from chromatic microlensing ($\sim1$-$1.5$ days). Our lens models converge to an Einstein radius of $\theta_E=(0.168^{+0.009}_{-0.005})\arcsec$, the smallest yet seen in a lensed SN system. The ``standard candle'' nature of \lensedsn provides magnification estimates independent of the lens modeling that are brighter than predicted by $\sim1.7^{+0.8}_{-0.6}$\,mag and $\sim0.9^{+0.8}_{-0.6}$\,mag for two of the four images, suggesting significant microlensing and/or additional substructure beyond the flexibility of our image-position mass models.
\end{abstract}

\section{Introduction}
\label{sec:intro}
Strong-gravitationally lensed supernovae (SNe) are rare events. In the strong lensing phenomenon, multiple images of a background source appear as light propagating along different paths are focused by a foreground galaxy or galaxy cluster. This requires a chance alignment along the line of sight between us the observers, the background source, and the foreground galaxy. Depending on the relative geometrical and gravitational potential differences of each path, the SN images typically appear delayed by hours to months (for galaxy-scale lenses) or years (for cluster-scale lenses). 

Robust measurements of this ``time delay'' can constrain the Hubble constant ($H_0$) and the dark energy equation of state (e.g., $w$) in a single step  \citep[e.g.,][]{refsdal_possibility_1964,linder_lensing_2011,paraficz_gravitational_2009,treu_time_2016,birrer_time-delay_2022,treu_strong_2022}. Lensed SNe have several advantages relative to quasars, which have historically been used for this purpose \citep[e.g.,][]{vuissoz_cosmograil_2008,suyu_dissecting_2010,tewes_cosmograil_2013,bonvin_h0licow_2017,birrer_h0licow_2019,bonvin_cosmograil_2018,bonvin_cosmograil_2019,wong_h0licow_2020}: 1) SNe fade quickly, enabling predictive experiments on the delayed appearance of trailing images more accurate models of the lens and source, as the SN (or quasar) and host fluxes are otherwise highly blended \citep{ding_improved_2021}, 2) SNe have predictable light curves, simplifying time-delay measurements and enabling SN progenitor system constraints, 3) the impact of microlensing is somewhat mitigated including a small ($\sim0.1$ day) ``microlensing time delay'' \citep{tie_microlensing_2018,bonvin_impact_2019} and less pronounced chromatic effects \citep{goldstein_precise_2018,foxley-marrable_impact_2018,huber_strongly_2019}, though this can still be a significant source of uncertainty when time delays are $\lesssim1$~day \citep[e.g.,][]{goobar_iptf16geu_2017}, and 4) time delay measurements for lensed SNe require much shorter observing campaigns than lensed quasars.  

While the advantages of using SNe for time-delay cosmography relative to other probes have been well-documented \citep[e.g.,][]{refsdal_possibility_1964,kelly_multiple_2015,goobar_iptf16geu_2017, goldstein_precise_2018,huber_strongly_2019,pierel_turning_2019,suyu_holismokes_2020,pierel_projected_2021,rodney_gravitationally_2021}, these events have proved extremely difficult to detect.  Since the first multiply-imaged SN discovery by \citet{kelly_multiple_2015}, there have been only four more such discoveries \citep{goobar_iptf16geu_2017,rodney_gravitationally_2021,chen_jwst-ers_2022,kelly_strongly_2022} despite dedicated surveys to increase the sample \citep[e.g.,][]{petrushevska_high-redshift_2016,petrushevska_searching_2018,fremling_zwicky_2020,craig_targeted_2021}. 

SNe of Type Ia (SNe\,Ia), those employed for decades as ``standardizable candles'' to measure cosmological parameters by way of luminosity distances and the cosmic distance ladder \citep[e.g.,][]{garnavich_supernova_1998,riess_observational_1998,perlmutter_measurements_1999,scolnic_complete_2018,brout_pantheon_2022}, are particularly valuable when strongly lensed. In addition to having a well-understood model of light curve evolution \citep{hsiao_k_2007,guy_supernova_2010,saunders_snemo_2018,leget_sugar_2020,kenworthy_salt3_2021,pierel_salt3-nir_2022}, their standardizable absolute brightness can provide additional leverage for lens modeling by limiting the uncertainty caused by the mass-sheet degeneracy \citep{falco_model_1985,kolatt_gravitational_1998,holz_seeing_2001,oguri_gravitational_2003,patel_three_2014,nordin_lensed_2014,rodney_illuminating_2015,xu_lens_2016,birrer_hubble_2022}, though only in cases where millilensing and microlensing are not extreme \citep[see][]{goobar_iptf16geu_2017,foxley-marrable_impact_2018,dhawan_magnification_2019}. The first such discovery was iPTF16geu \citep{goobar_iptf16geu_2017}, which had image separations resolved using adaptive-optics (AO) and \hst, with very short time delays of $\sim0.25$-$1.5$ days \citep{dhawan_magnification_2019}. Nevertheless, the detection and analysis of objects like iPTF16geu are critical to the future of lensed SN research as unresolved, galaxy-scale lenses are expected to be relatively common amongst lensed SN discoveries made with the Vera C. Rubin Observatory \citep{collett_population_2015,goldstein_rates_2019,wojtak_magnified_2019}.

LensWatch\footnote{\url{https://www.lenswatch.org}} is a collaboration with the goal of finding gravitationally lensed SNe, both by monitoring active transient surveys \citep[e.g.,][]{fremling_zwicky_2020,jones_young_2021} and by way of targeted surveys \citep{craig_targeted_2021}. The collaboration maintains a Cycle 28 \textit{HST} program\footnote{\href{https://archive.stsci.edu/proposal_search.php?id=16264&mission=hst}{\textit{HST}-GO-16264}}, given long-term (3-cycle) target of opportunity (ToO) status due to the relatively low expected lensed SN rates. The program includes three ToO triggers (two non-disruptive, one disruptive), and was designed to provide the high-resolution follow-up imaging for a ground-based lensed SN discovery, which is critical for galaxy-scale multiply-imaged SNe due to their small image separations \citep[e.g.,][]{goobar_iptf16geu_2017}. 

A new multiply-imaged SN\,Ia was discovered in 2022 August by the Zwicky Transient Facility \citep[ZTF;][]{fremling_zwicky_2020}\footnote{\url{https://www.wis-tns.org/object/2022qmx}}, subsequently classified and analyzed by \citet[][hereafter G22]{goobar_sn_2022}. The separate four images of this SN 2022qmx (aka ``SN Zwicky'') were spatially unresolved in ground-based imaging with separations of $\lesssim0.3\arcsec$. In order to provide reliable photometry and the data necessary for accurate lens modeling of the system, optical space-based observations are ideal. We therefore report the first observations and results of the LensWatch collaboration, which triggered \textit{HST} GO program 16264 to schedule follow-up imaging of \lensedsn. 

This work is the first in a series of papers that utilize data from the LensWatch program. Section \ref{sec:obs} gives an overview of \lensedsn and presents the final \hst observation characteristics including triggering, orbit design, and implementation. Section \ref{sec:lens_model} details our lens modeling methodology and constraints on the lensing system, and our analysis of \lensedsn (including photometry and measurements of time delays and magnifications) are reported in Section \ref{sec:zwicky_analysis}. We conclude with a discussion of implications of this new dataset, as well as future observation plans, in Section \ref{sec:conclusion}. 

\section{Observing with \textit{HST}}
\label{sec:obs}
 As possible discovered lensed system configurations are highly variable, it is necessary to design a custom follow-up campaign for each new discovery. We therefore give an overview of the lensing system and SN characteristics for \lensedsn, and then the subsequent observational choices made for the LensWatch \hst ToO trigger.

\subsection{The Multiply-Imaged \lensedsn}
\label{sub:zwicky}
The discovery, description of ground-based observations, and initial analysis of \lensedsn are presented by G22. Briefly, the SN was discovered by ZTF at Palomar Observatory under the Bright Transient Survey \citep[BTS;][]{fremling_zwicky_2020} on August 1, 2022 (MJD $59792$). The SED Machine (SEDM) and Nordic Optical Telescope (NOT) provided spectroscopic classification of \lensedsn as a Type Ia (SN\,Ia) at $z=0.35$ and near maximum light on August 21-22, 2022 (MJD $59812$-$59813$). Although the multiple images were not resolved by ZTF, the inferred absolute magnitude of \lensedsn for this redshift was $\sim3$ magnitudes brighter than normal, suggesting the presence of strong gravitational lensing. G22 also obtained subsequent spectroscopic observations from the Keck observatory, Hobby-Eberly Telescope, and the Very Large Telescope (VLT), which led to a final SN redshift of $z=\snz$ and lensing galaxy redshift of $z=\lensz$. The multiple images of \lensedsn were first resolved with the Keck telescope Laser Guide Star aided Adaptive Optics (LGSAO) Near-IR Camera 2 (NIRC2) on September 15, 2022 (MJD $59837$; see G22 for details).

\subsection{ToO: Filter Choices \& Orbit Design}
\label{sub:orbit}
Roughly $12$ days after the spectroscopic classification of \lensedsn, we used a non-disruptive \hst ToO trigger to obtain follow-up WFC3/UVIS and IR images of the lensing system. The average turnaround for a non-disruptive ToO trigger is $\gtrsim21$ days, but close coordination with the \textit{HST} scheduling team at STScI led to receiving our first images after $17$ days on September 21, 2022 (MJD $59843$), or $\sim44$ observer-frame ($\sim32$ rest-frame) days post-discovery and $\sim37$ observer-frame ($\sim27$ rest-frame) days after maximum brightness. 

\begin{figure}
    \centering
    \includegraphics[width=\columnwidth]{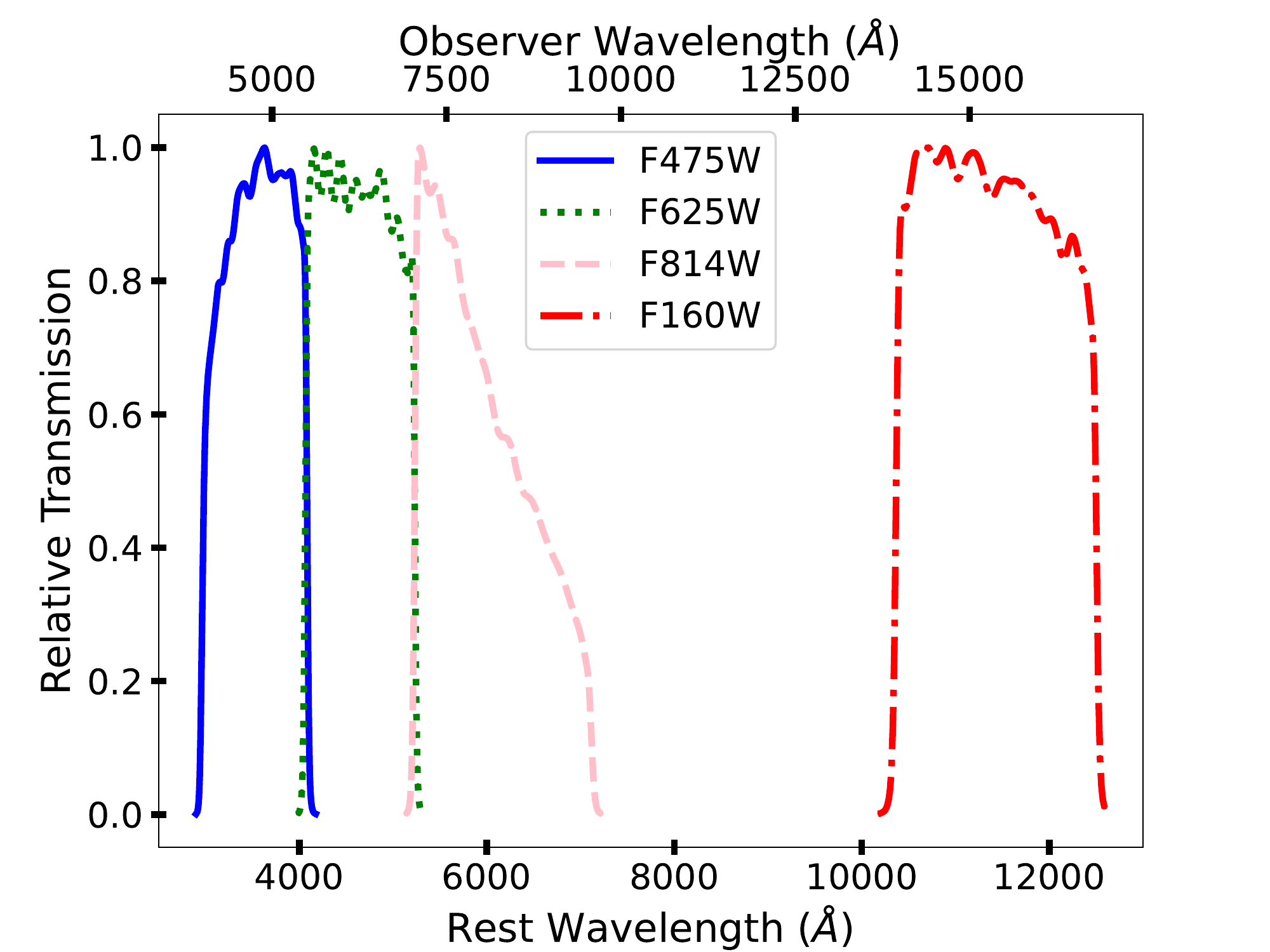}
    \caption{The \hst filters used to observe \lensedsn, with rest-frame wavelength on the lower axis and observer-frame wavelength of the upper axis. The three bluer filters are from WFC3/UVIS, while F160W is from WFC3/IR.}
    \label{fig:filters}
\end{figure}

The anticipated image separations for a galaxy-scale lens of this mass and redshift are small enough that resolving the individual images with WFC3/IR ($0.13\arcsec$/pix), where the point-spread function (PSF) is severely undersampled, is unlikely. For the purposes of accurate photometry and lens modeling, the highest possible resolution imaging is required, and we therefore turned to WFC3/UVIS ($0.04\arcsec$/pix) to resolve the multiple images. We selected the F814W, F625W, and F475W filters to provide non-overlapping coverage across the full optical wavelength range ($\sim$ 3,500-6,000~\angstrom in the rest-frame; see Figure \ref{fig:filters} and Table \ref{tab:filters}). Additionally, the ground-based follow-up campaign of \lensedsn included (resolved) H-band Keck-AO imaging, and we therefore included (unresolved) WFC3/IR F160W observations to provide overall calibration and extra information about the lensing system. 

\begin{figure*}[t!]
    \centering
    \includegraphics[width=\linewidth]{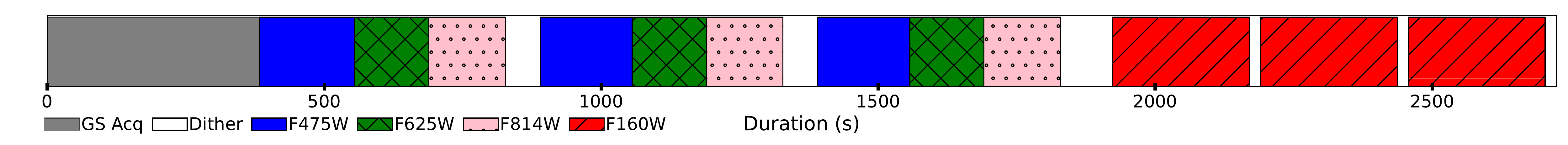}
    \caption{The layout of the orbit used for these \hst observations. The dither sections (white) include other overheads such as filter changing.}
    \label{fig:orbit}
\end{figure*}

\begin{figure*}
    \centering
    \includegraphics[trim={0cm 9.35cm 9.5cm 0cm},clip,width=\linewidth]{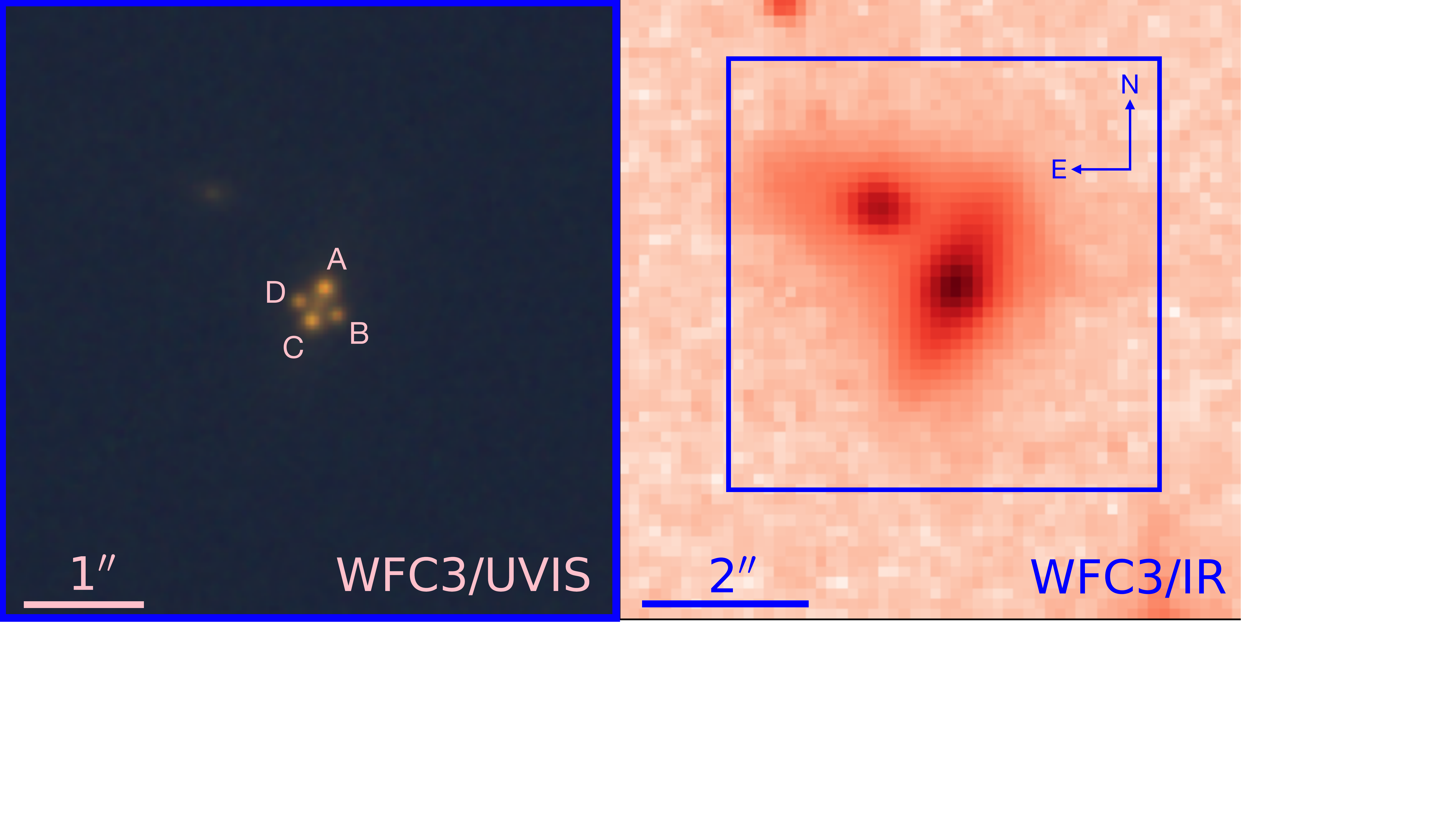}
    \caption{ (Left) A WFC3/UVIS combined color image (R=F814W, G=F625W, B=F475W) of \lensedsn, with the multiple images labeled as A-D. (Right) WFC3/IR F160W image of the galaxy lensing of \lensedsn, where the multiple images are not resolved. The orientation of the SN host galaxy (northeast of the lens) suggests there is not much intervening matter along the line of sight apart from the lensing galaxy. The blue square is the footprint of the WFC3/UVIS image. }
    \label{fig:color_im}
\end{figure*}

The four filters were efficiently packed into a single orbit of observing using the $512\times512$ subarrays for both UVIS and IR imaging, even with three dithers per filter to reduce the impact of cosmic rays and provide optimal sampling of the (Figure \ref{fig:orbit}). The four images of \lensedsn were successfully resolved in the three UVIS filters, which provided a full-color image (Figure \ref{fig:color_im}).

\begin{table}[h]
\caption{\label{tab:filters}\hst WFC3 photometric filter definitions and exposure times.}
\begin{tabular*}{\linewidth}{@{\extracolsep{\stretch{1}}}*{5}{c}}
\toprule
 Band&Rest $\lambda_{\rm{eff}}$ &Obs $\lambda_{\rm{eff}}$&Instrument&Exp. Time\\
 &$(\angstrom)$ &$(\angstrom)$&&(s)\\
\hline
F475W&$3,549$&$4,792$&UVIS&$126$\\
F625W&$4,636$&$6,258$&UVIS&$39$\\
F814W&$5,965$&$8,053$&UVIS&$60$\\
F160W&$11,40$2&$15,392$&IR&$207$\\

\end{tabular*}
\end{table}

\section{Modeling of the Lensing System}
\label{sec:lens_model}
\subsection{Analysis Methods}
\label{sub:lens_model}
In this section, we summarize the lens modeling analysis we carried out using the \hst data presented in Section \ref{sec:obs}, leading to insights into the lensing galaxy mass (see Appendix). Given the very low number of identified strongly lensed SNe, this procedure has mainly been applied to strongly lensed quasars, e.g., by the $H_0$ Lenses in COSMOGRAIL's Wellspring (H0LiCOW) collaboration \citep[e.g.,][]{wong_h0licow_2020}. 

For galaxy-scale lenses, the lens mass distribution is usually described by profiles such as the singular isothermal ellipsoid \citep[SIE;][]{kormann_isothermal_1994} or the singular power-law elliptical mass distribution \citep[SPEMD;][]{barkana_fast_1998}, in combination with an external shear component describing the influence from massive line-of-sight objects \citep{mccully_quantifying_2017}. These parameters can be constrained by the observed image positions alone (those measured in Section \ref{sub:phot}), and/or through the pixel intensities of the \textit{HST} images. This requires a model of the lens light distribution, which is typically described by one or more stacked \sersic profiles \citep{de_vaucouleurs_sur_1948, sersic_influence_1963}, as well as a model for the lensed SN represented by a PSF (described in Section \ref{sub:phot}). 

Given these different potential methodologies, we used three independent software packages and five total methods to carry out independent analyses of \lensedsn, which provides an examination of potential modeling systematics and allows us to marginalize over them \citep[e.g.,][]{ertl_tdcosmo_2022,shajib_tdcosmo_2022}. The three software packages are \yscode \citep{shu_lfit_2016}, \lenstronomy \citep{birrer_gravitational_2015,birrer_lenstronomy_2018,birrer_lenstronomy_2021} and the Gravitational Lens Efficient Explorer \citep[\GLEE;][]{suyu_halos_2010,suyu_disentangling_2012,ertl_tdcosmo_2022}, and the five methods explored here are summarized by Table \ref{tab:lens_models}. 


Using the first four models reported in Table \ref{tab:lens_models}, we found a significantly higher reduced-$\chi^2$ when fitting both image positions and fluxes compared to fitting only image positions, indicating the presence of substructure and/or microlensing not captured by the lens modeling process. Additionally, the SALT+LS modeling method first performs PSF photometry to obtain individual image fluxes, which are fit directly with the SALT2.4 model \citep{betoule_improved_2014}. This initial fit is used to provide a prior on the magnifications given the standardizability of \lensedsn as an SN\,Ia (see Section \ref{sub:sn_results} for details), then the lens model parameters are constrained with the image positions. As is apparent in the following section, the SALT+LS model magnification estimates agree with the models constrained with positions only for images B and D but not A and C, further supporting our assumption that images A and C are impacted by additional factors.

As a result of the above, we rely only on the positions of the multiple images to constrain the initial four lens models. Our interpretation is that, pending updated difference image photometry, there is some combination of additional microlensing and/or millilensing impacting images A and C. We discuss this more in Section \ref{sec:conclusion}, and will wait for the upcoming template image to improve both our lens models and photometry. For the remainder of this work, we refer to the models used to constrain the lensing system parameters (\yscode, LS1, LS2, \GLEE) as the ``primary'' models, and we refer to the SALT+LS model by name.


\begin{table*}
    \centering
    \caption{\label{tab:lens_models}Summary of lens modeling methodologies.}
    
    \begin{tabular*}{\linewidth}{@{\extracolsep{\stretch{1}}}*{5}{c}}
\toprule
Name&Code&Lens Model Components&Fitted Filters&Modeling team\\
\hline
    
    \yscode &\yscode &SIE&All&YS\\
    LS1&\lenstronomy&SIE&All&NA, AJS\\
    LS2&\lenstronomy&Power-law+$\gamma_{\rm ext}$&All&LM, SB\\
    \GLEE&\GLEE&Power-law+$\gamma_{\rm ext}$&F814W&SE, SS, SHS\\ 
    \hline
    SALT+LS$^a$&\lenstronomy&Power-law+$\gamma_{\rm ext}$&All&XH, WS, ES, SA\\

    \end{tabular*}
    
    \noindent $^a$ The SALT+LS model also includes constraints from the Type Ia absolute magnitude (see Section \ref{sub:lens_model}).

\end{table*}

\subsection{Lens Model Constraints}
\label{sub:lens_results}
\begin{figure}
    \centering
    \includegraphics[trim={0cm 0cm .5cm 1.6cm},clip,width=\columnwidth]{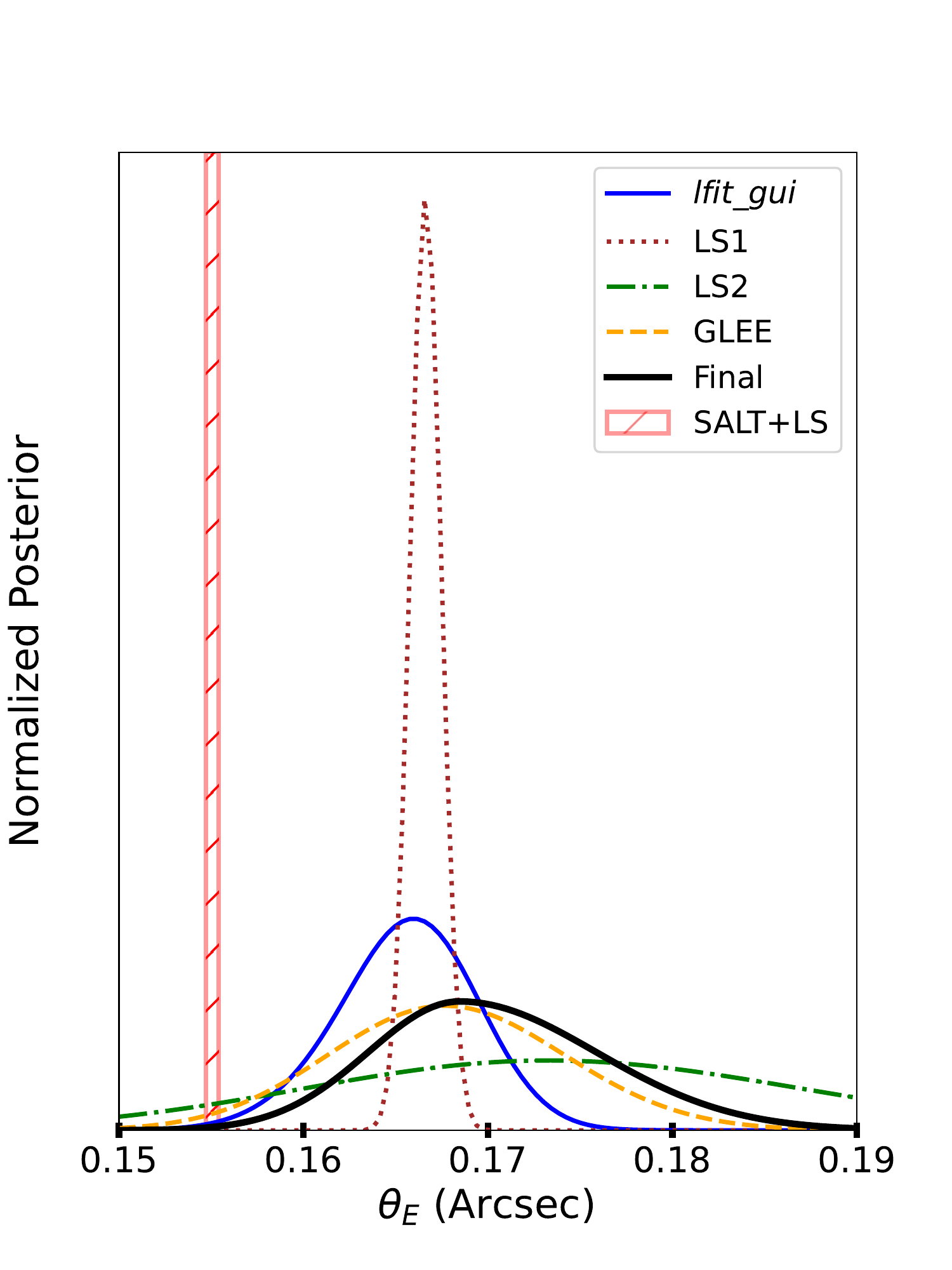}
    \caption{Normalized posterior distributions for $\theta_E$ of the four primary lens models, and their combined constraint (dark grey mean, light grey uncertainty). The SALT+LS model constraint is also shown for comparison (red slashed).}
    \label{fig:lens_thetae}
\end{figure}

\begin{figure}
    \centering
    \includegraphics[trim={1.5cm 1.25cm 1.5cm 2.25cm},clip,width=\columnwidth]{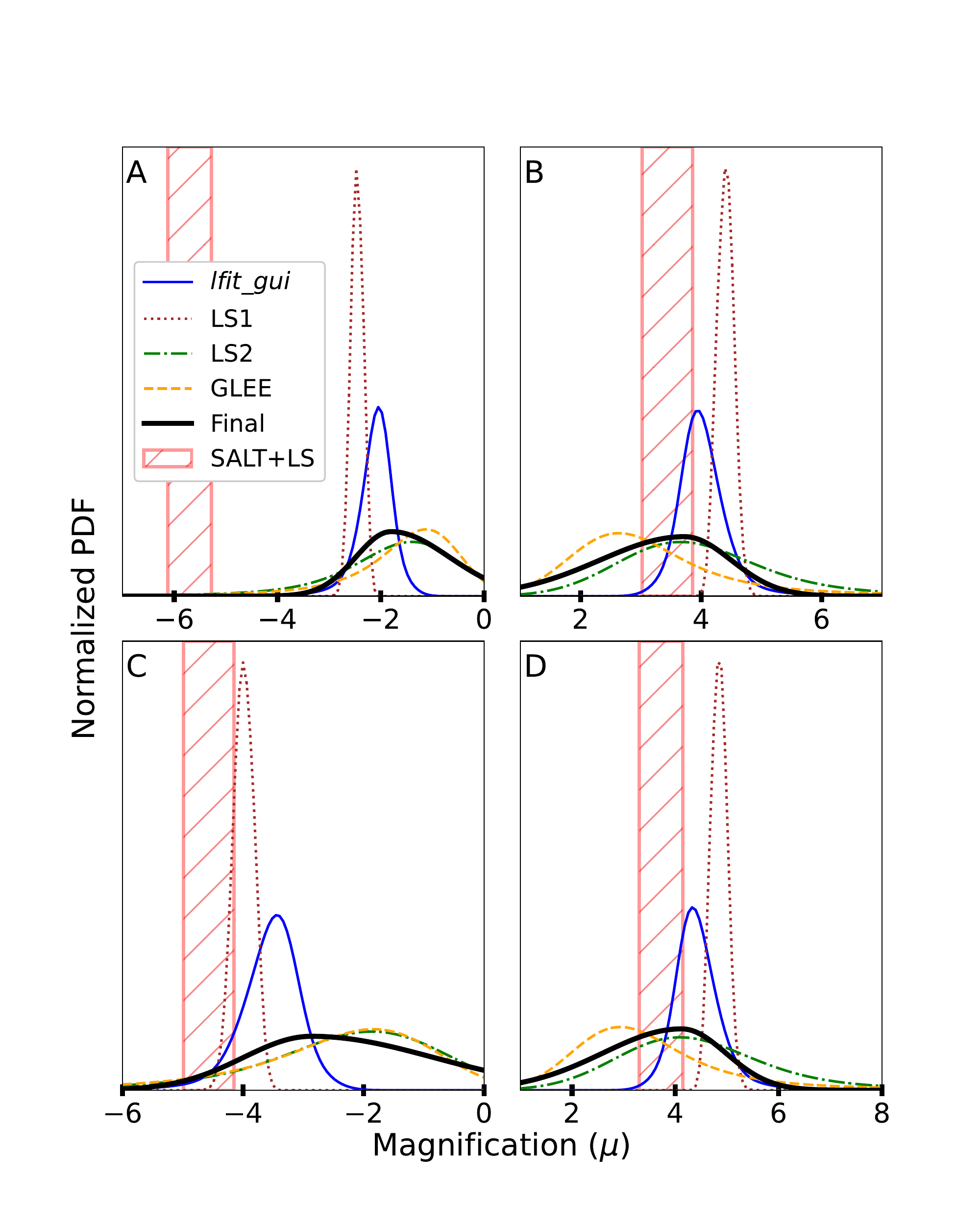}
    \caption{Normalized posterior distributions of the four primary lens models for the four image magnifications, and their combined constraints (dark grey mean, light grey uncertainty). The SALT+LS model constraints are also shown for comparison (red slashed). Note that images A,C have negative magnification and B,D have positive magnification, indicating all lens models agree on the parity of each image in addition to the absolute value of the magnification.}
    \label{fig:lens_mu}
\end{figure}

\begin{figure}
    \centering
    \includegraphics[trim={0.75cm 2cm 1.5cm 4cm},clip,width=\columnwidth]{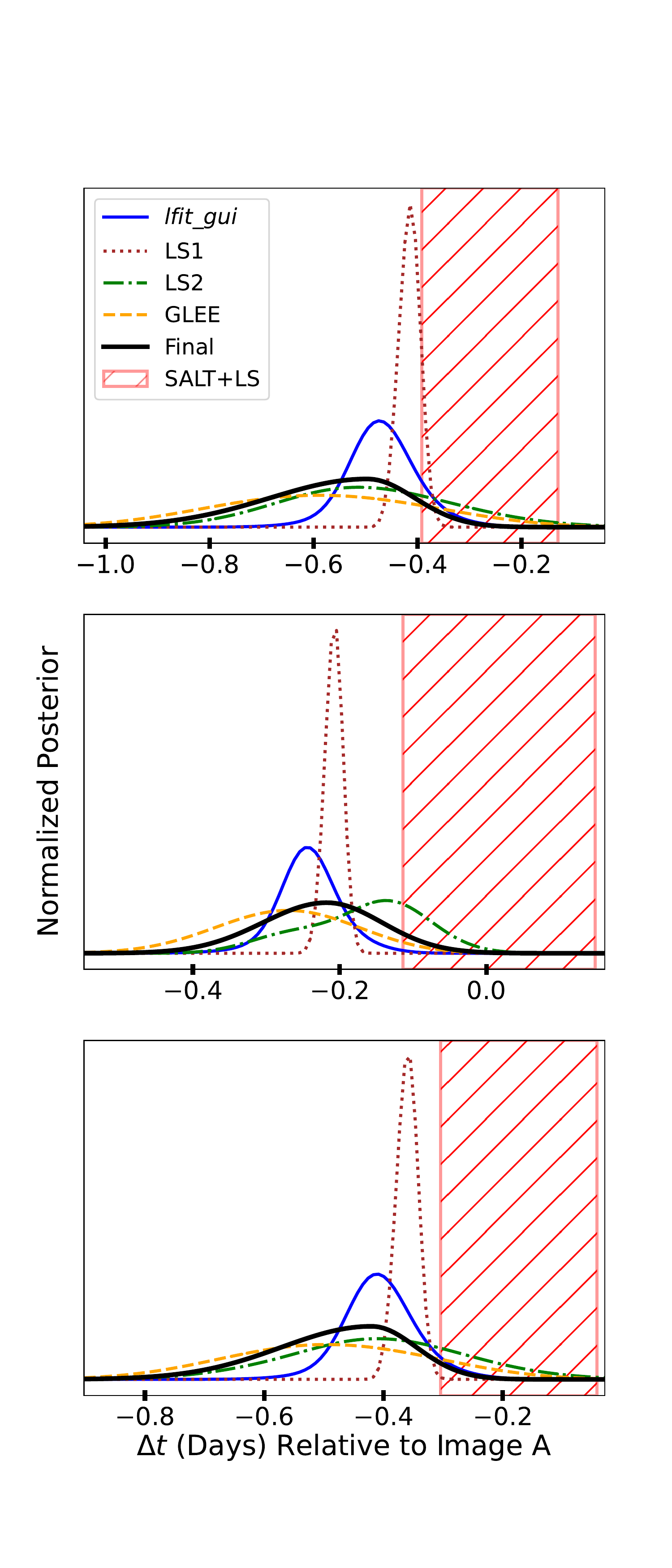}
    \caption{Normalized posterior distributions of the four primary lens models for the time delays relative to image A, and their combined constraints (dark grey mean, light grey uncertainty). The SALT+LS model constraints are also shown for comparison (red slashed).}
    \label{fig:lens_dt}
\end{figure}

Each of the primary lens modeling methods described in Section \ref{sub:lens_model} was used to independently constrain the magnifications and time delays for each SN image, as well as the Einstein radius ($\theta_E$) of the lensing galaxy. These results are summarized in Table \ref{tab:lens_results}, where $\Delta t_{iA}$ refers to the relative delay between the $i^{\rm{th}}$ image and image A (i.e., $t_i-t_A$), and we also report the ``Final'' combined measurement for each parameter. This final value is the equally-weighted average of each lens modeling result, and the uncertainty is a combination of the standard deviation of the primary model results and the statistical scatter in the average posterior distribution. As the models are equally weighted to obtain our final values for each parameter, we show normalized posterior distributions simply for visual comparison in Figures \ref{fig:lens_thetae}-\ref{fig:lens_dt}. Note that the models from Table \ref{tab:lens_models} with fewer parameters also have narrower posterior distributions. We also include the results of the SALT+LS model, which reveals the impact of including information about the SN\,Ia standardizability. The additional specific lensing model parameters measured by each method, as well as more details about the modeling processes, are given in the Appendix. 

While we expect improved constraints following the LensWatch template image scheduled for 2023 as we will be able to disentangle the SN and lensing galaxy flux more reliably, the level of agreement between the primary modeling methods gives us confidence in the final constraints and uncertainties. We also find good agreement between these key parameters with the modeling of G22, which used only the resolved near-IR Keck data. We note that our measured Einstein radius of $\theta_E=(0.168^{+0.009}_{-0.005})\arcsec$ is the smallest detected value for a multiply-imaged SN thus far, and corresponds to a lens mass of $\sim8\times10^9M_\odot$. The similar lensed SN iPTF16geu had a measured $\theta_E=0.29\arcsec$ \citep{more_interpreting_2017,mortsell_lens_2020}, with time delays of $\sim1$ day \citep{dhawan_magnification_2019}. Here we predict time delays of $\sim0.2$-$0.5$ days for each of the images of \lensedsn, which is well below the predicted time-delay measurement uncertainty for even a resolved and well-sampled lensed SN\,Ia due to the impacts of microlensing \citep[e.g.,][]{pierel_projected_2021,huber_holismokes_2022}. 

We also note that the SALT+LS method, which uniquely uses the measured photometry to infer a standardized absolute magnitude measurement of \lensedsn and sets a prior on the image magnifications (see Section \ref{sub:lens_model}), is generally in good agreement with other methods apart from the predicted magnifications for images A and D and a slightly lower $\theta_E$. The method also significantly reduced the plausible model parameter space (see Appendix), which lends weight to the claims that SN\,Ia standardization can significantly improve lens modeling efforts when microlensing is minimal. By implementing models that did not include this extra step alongside SALT+LS, the relative agreement (or disagreement) between methods was a useful indicator of additional substructure/microlensing beyond the primary lens modeling flexibility.

\begin{table*}
    \centering
    \caption{Lens modeling constraints on key parameters.}
    \label{tab:lens_results}
    \begin{tabular*}{\linewidth}{@{\extracolsep{\stretch{1}}}*{7}{c}|c}
\toprule
Parameter&Unit&\yscode&LS1&LS2&\GLEE&\textbf{Final}&SALT+LS\\
\hline
$\theta_E$&$\arcsec$&$0.166^{+0.0010}_{-0.0019}$&$0.167^{+0.0005}_{-0.0005}$&$0.173^{+0.0086}_{-0.0071}$&$0.168^{+0.0046}_{-0.0037}$&$\mathbf{0.168^{+0.009}_{-0.005}}$&$0.155\pm0.0004$\\
\hline
$^a\mu_A$&&$-2.05^{+0.17}_{-0.26}$&$-2.46^{+0.12}_{-0.11}$&$-1.45^{+0.49}_{-1.05}$&$-1.26^{+0.56}_{-1.43}$&$\mathbf{-1.81^{+0.90}_{-0.89}}$&$-5.70\pm0.42$\\
$\mu_B$&&$3.96^{+0.32}_{-0.20}$&$4.41^{+0.12}_{-0.14}$&$3.75^{+1.17}_{-0.65}$&$2.78^{+1.71}_{-0.67}$&$\mathbf{3.72^{+1.04}_{-1.24}}$&$3.44\pm0.42$\\
$\mu_C$&&$-3.47^{+0.25}_{-0.43}$&$-3.99^{+0.16}_{-0.16}$&$-1.94^{+0.67}_{-1.45}$&$-2.07^{+0.95}_{-2.29}$&$\mathbf{-2.87^{+1.51}_{-1.50}}$&$-4.57\pm0.42$\\
$\mu_D$&&$4.36^{+0.37}_{-0.23}$&$4.85^{+0.14}_{-0.15}$&$4.19^{+1.33}_{-0.78}$&$3.10^{+1.92}_{-0.73}$&$\mathbf{4.12^{+1.19}_{-1.36}}$&$3.72\pm0.42$\\
\hline
$\Delta t_{BA}$&Days&$-0.48^{+0.05}_{-0.03}$&$-0.41^{+0.01}_{-0.02}$&$-0.50^{+0.15}_{-0.12}$&$-0.59^{+0.22}_{-0.20}$&$\mathbf{-0.50^{+0.15}_{-0.21}}$&$-0.26\pm0.13$\\
$\Delta t_{CA}$&Days&$-0.24^{+0.03}_{-0.02}$&$-0.21^{+0.01}_{-0.01}$&$-0.15^{+0.04}_{-0.11}$&$-0.27^{+0.09}_{-0.09}$&$\mathbf{-0.22^{+0.10}_{-0.10}}$&$0.02\pm0.13$\\
$\Delta t_{DA}$&Days&$-0.41^{+0.04}_{-0.03}$&$-0.36^{+0.01}_{-0.02}$&$-0.41^{+0.12}_{-0.11}$&$-0.50^{+0.18}_{-0.16}$&$\mathbf{-0.42^{+0.12}_{-0.18}}$&$-0.17\pm0.13$\\

\hline
\hline
    \end{tabular*}
\begin{flushleft}
$^a$ See appendix for $\kappa$ and $\gamma$ results for each lens model.
\end{flushleft}
\end{table*}


\section{Analysis of \lensedsn}
\label{sec:zwicky_analysis}
\subsection{\hst Photometry}
\label{sub:phot}
Due to the compact nature of the lensing system and difficulty in disentangling the SN and lens galaxy flux, an identical ``template'' epoch has been scheduled for $\sim6$-$12$ months after the first, once the SN has long faded. This will provide more precise measurements and constraints for \lensedsn and the lensing system in general. In the meantime, we have used PSF photometry to optimally measure the brightness of each SN image.

\hst photometry for \lensedsn was measured using PSF photometry on the WFC3/UVIS ``FLC'' images, which are individual exposures that have been bias-subtracted, dark-subtracted, and flat-fielded but not yet corrected for geometric distortion. The UVIS data processing also includes a charge transfer efficiency (CTE) correction, which results in FLC images instead of the FLT images used for WFC3/IR. The WFC3/UVIS2 pixel area map (PAM) for the corresponding subarray was also applied to each exposure to correct for pixel area variations across the images\footnote{\href{https://www.stsci.edu/hst/instrumentation/wfc3/data-analysis/pixel-area-maps}{https://www.stsci.edu/hst/instrumentation/wfc3/data-analysis/pixel-area-maps}}. 

In most cases, the individual exposures for each filter are ``drizzled'' together to create a single image (e.g., Figure \ref{fig:color_im}). Here, we primarily are concerned with precisely measuring the position and brightness of each SN image, which (without a template image) requires accurate fitting of a PSF model. Drizzled images can introduce inconsistencies into the modeling of a PSF, and so we restrict ourselves to the FLCs to preserve the PSF structure. We use the standard \hst PSF models\footnote{\url{https://www.stsci.edu/hst/instrumentation/wfc3/data-analysis/psf}} to represent the PSF, which also take into account spatial variation across the detector.

For each UVIS filter, we implement a Bayesian nested sampling routine\footnote{\textsc{nestle}: \url{http://kylebarbary.com/nestle}} to simultaneously constrain the (common) SN flux and relative position in all three FLCs for all four SN images. Each PSF was fit to the multiple SN images within a $5\times5$ pixel square in an attempt to limit the contamination of both the lensing galaxy (as we assume a constant background in the fitted region) and other SN images. The PSF full-width at half-maximum (FWHM) for WFC3/UVIS is $<2$ pixels, so this PSF size should include $\sim99\%$ of the total SN flux and not be contaminated by significant flux from the other images (each $\gtrsim5$ pixels away). 

The final measured flux is the integral of each full fitted PSF model, which is $101\times101$ pixels and large enough to approximately contain all of the SN flux. These corrected fluxes were converted to AB magnitudes using the time-dependent inverse sensitivity and filter pivot wavelengths provided with each data file. The final measured magnitudes and colors are reported in Table \ref{tab:im_mags}. 

\begin{table*}
    \centering
    \caption{\label{tab:im_mags}Photometry and colors measured for each image of \lensedsn in AB magnitudes.}
    
    \begin{tabular*}{\linewidth}{@{\extracolsep{\stretch{1}}}*{7}{c}}
\toprule
Image&\multicolumn{1}{c}{F475W}&\multicolumn{1}{c}{F625W}&
\multicolumn{1}{c}{F814W}&\multicolumn{1}{c}{F475W$-$F625W}&\multicolumn{1}{c}{F475W$-$F814W}&
\multicolumn{1}{c}{F625W$-$F814W}\\
\hline
A&23.22$\pm$0.04&21.67$\pm$0.02&20.67$\pm$0.01&1.55$\pm$0.04&2.55$\pm$0.04&1.00$\pm$0.02\\
B&24.31$\pm$0.07&22.65$\pm$0.03&21.71$\pm$0.02&1.66$\pm$0.08&2.60$\pm$0.08&0.94$\pm$0.04\\
C&23.35$\pm$0.04&21.90$\pm$0.02&20.88$\pm$0.02&1.44$\pm$0.04&2.47$\pm$0.04&1.02$\pm$0.03\\
D&24.26$\pm$0.07&22.72$\pm$0.04&21.60$\pm$0.02&1.53$\pm$0.08&2.65$\pm$0.07&1.12$\pm$0.04\\
\hline
    \end{tabular*}

\end{table*}

\begin{figure}
    \centering
    \includegraphics[trim={4.5cm 1.5cm 4.5cm .5cm},clip,width=\columnwidth]{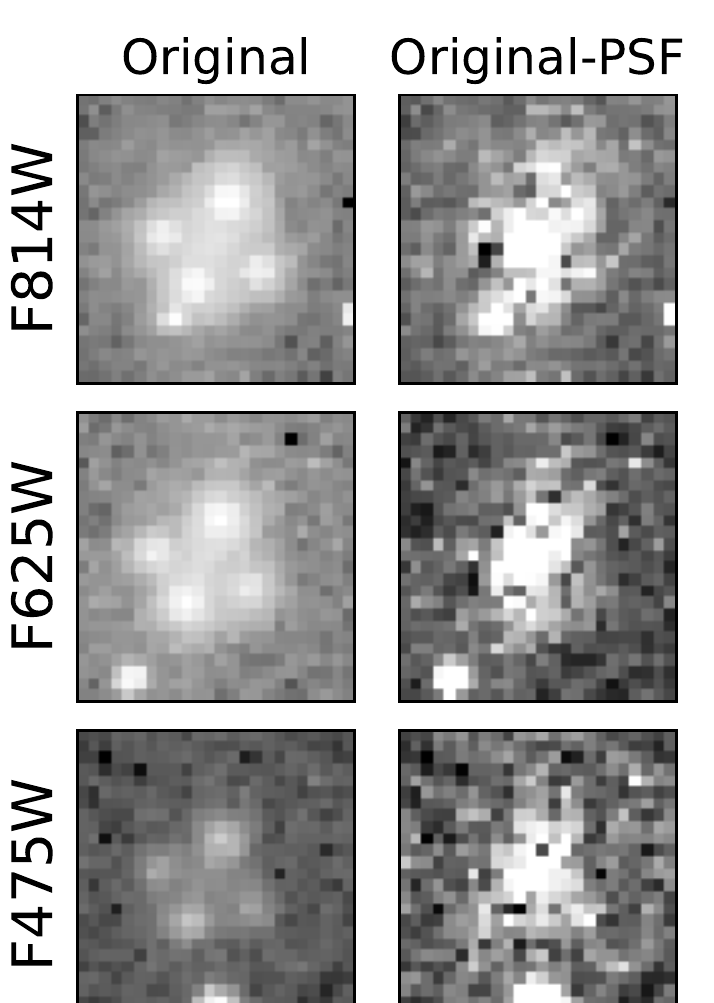}
    \caption{Results (right) of subtracting the best-fit PSF models from a single FLC for each WFC3/UVIS filter (left). Some residuals remain, particularly for the brightest images, but our planned template epoch will significantly improve the measured fluxes.}
    \label{fig:psf_sub}
\end{figure}

\subsection{Single Epoch Time Delays and Magnifications from \hst Photometry}
\label{sub:sn_results}
We use the single epoch of \textit{HST} photometry from Table \ref{tab:im_mags} to constrain the time delays and magnifications for the multiple images of \lensedsn in the manner of \citet{rodney_gravitationally_2021}. As measuring the difference in time of peak brightness for each image directly \citep[e.g.,][]{rodney_sn_2016} is not possible with a single epoch, we instead constrain the age of each SN image given a single light curve model. The relative age difference for each image is also a measure of the time delay, though we note this method is only possible because we have a reliable model for the light (and color) curve evolution as \lensedsn is of Type Ia. As we use some information from the unresolved light curve in G22, we also fit the data with the commonly used Spectral Adaptive Lightcurve Template \citep[SALT2;][]{guy_supernova_2010}, which provides a simple parameterization of the Type Ia SN normalization ($x_0$), shape or stretch ($x_1$), and color ($c$) used for light curve standardization. The remaining SALT2 parameters are a time of peak B-band brightness ($t_{pk}$) and the SN redshift. 

We fit the photometry of the multiple images simultaneously using the \sntd software package \citep{pierel_turning_2019}, where we also include the known effects of Milky Way dust ($E(B-V)=0.16, \ R_V=3.1$) based on the maps of \citet{schlafly_measuring_2011} and extinction curve of \citet{fitzpatrick_correcting_1999}. Additionally, we use the simulations of \citet{goldstein_precise_2018} to estimate the additional uncertainty introduced by chromatic microlensing. Using the time of peak estimate from G22 our observations are $\sim46$ days post explosion, which corresponds to $\sim0.05, \ 0.05$, and $0.11\rm{mag}$ of additional color uncertainty in rest-frame $U-B$ ($\sim$F475W$-$F625W), $B-V$ ($\sim$F625W$-$F814W), and $U-V$ ($\sim$F475W$-$F814W) respectively \citep[95\% confidence; see][figure 5]{goldstein_precise_2018}. We add these uncertainties in quadrature to the color uncertainties shown in Table \ref{tab:im_mags} for the fitting process. 

We follow the methods outlined by \citet{rodney_gravitationally_2021} to measure time delays for \lensedsn, which performed a similar analysis with the single epoch of SN Requiem. This process uses the SN color curves to constrain the time delay (with the \sntd ``Color'' method), and then fits for relative magnifications (with the \sntd ``Series'' method). Unlike the analysis of SN Requiem, an unresolved light curve exists for \lensedsn and in G22 was analyzed to give $t_{pk}=\snpeak, \ c=\snc, \ x_1=\snstretch$. While our single epoch of photometry should constrain the color parameter, it will be unable to constrain the $x_1$ parameter and there will be significant degeneracies between time delays and $t_{pk}$ \citep[as seen in][]{rodney_gravitationally_2021}. We therefore allow the $t_{pk}$ parameter, which here describes the time of peak for image A (see Figure \ref{fig:color_im} for naming convention), to vary only within fifteen days of \snpeak. We also fix $x_1$ to the parameter derived by G22 ($x_1=\snstretch$), mainly to ensure an accurate light curve standardization. We repeated the fitting first following the choice in \citet{rodney_gravitationally_2021} to set $x_1=0$ and second allowing $x_1$ to vary within 3$\sigma$ of the value measured in G22, and found these varied the time delays by $\lesssim0.5$ days, well within our measurement error bars. We also checked the difference in measured time delays when fixing the value of $t_{pk}$ to \snpeak and found a difference of $\lesssim0.5$ days. Finally, we note that the additional uncertainty added due to chromatic microlensing changes the measured time delay by $\lesssim0.5$ days as well, but increases the measurement uncertainties by $\sim1$-$1.5$ days.

\begin{figure}
    \centering
    \includegraphics[width=\linewidth]{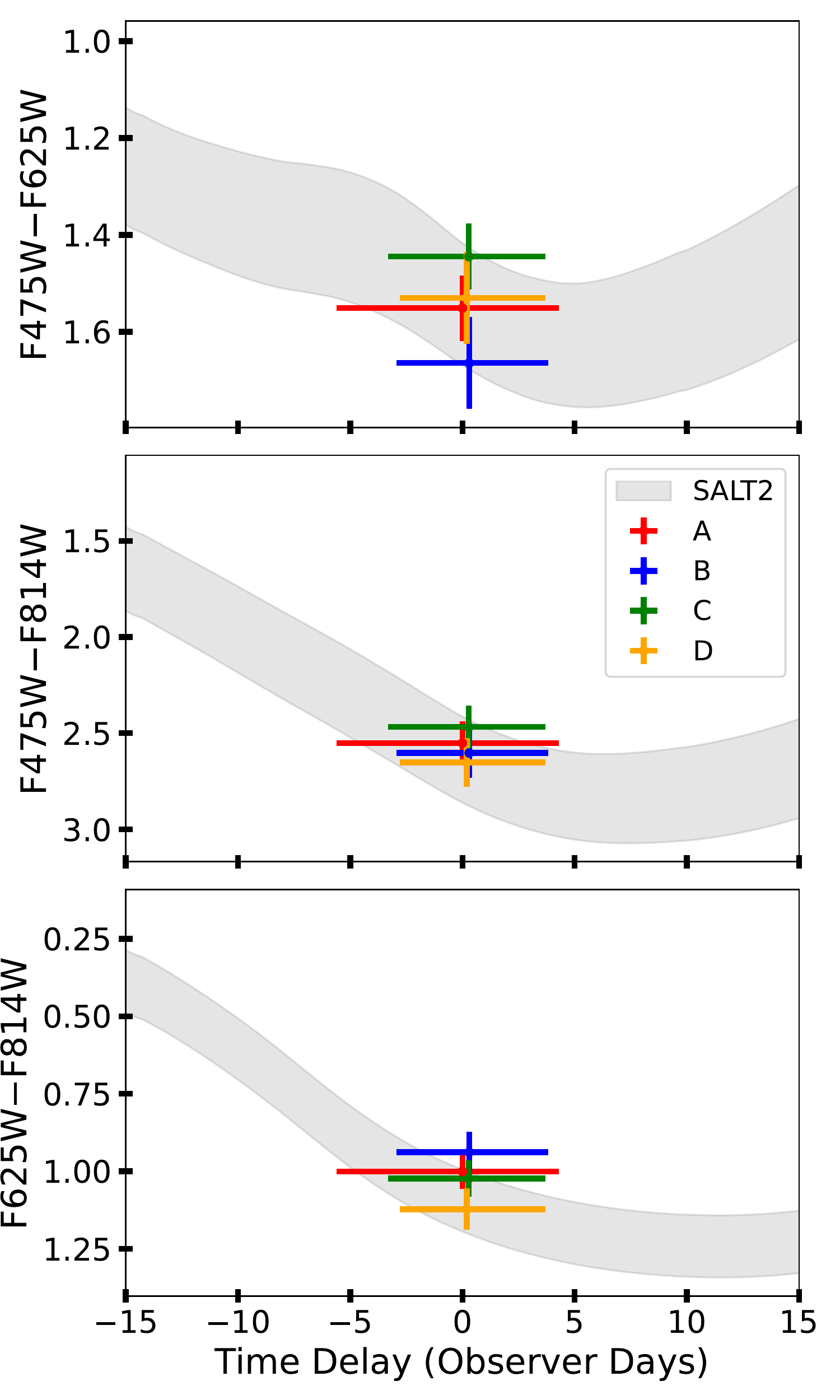}
    \caption{Measurements of time delay and color for each image of \lensedsn, with image A the reference image (i.e., $\Delta t=0$). The vertical error bars are the photometric precision based on the work in Section \ref{sub:phot} combined with an additional microlensing uncertainty}, while the horizontal error bars are the $16^{\rm{th}}$ and $84^{\rm{th}}$ quantiles for the time delay posterior of images B-D (columns 2 and 3-5 in Figure \ref{fig:sntd_corner}), and $t_{pk}$ for image A. The grey shaded region is the best-fit SALT2 model from the \sntd Color method.
    \label{fig:color_curves}
\end{figure}

\sntd finds a common value for $c$ amongst all \lensedsn images while varying the time delays of images B-D relative to the value of $t_{pk}$, which describes image A, within relative bounds of $[-15,15]$ days. Figure \ref{fig:color_curves} shows the measured colors and time delays for \lensedsn with the best-fit SALT2 model overlaid. While all colors are used in the fit simultaneously, the photometric and model uncertainties mean F625W-F814W provides the most constraining power, followed by the colors that include the rest-frame ultraviolet. After these time delays have been measured with the \sntd Color method, we fix all best-fit parameters and use the \sntd Series method to estimate magnification ratios for images B-D (within bounds of $[0.01,100]$) relative to image A. The fitting procedure for the light curve parameters is summarized in Table \ref{tab:fitting}. 

\begin{table*}
    \centering
    \caption{\label{tab:fitting}Summary of the SALT2 parameters used in fitting \lensedsn.}
    
    \begin{tabular*}{\linewidth}{@{\extracolsep{\stretch{1}}}*{5}{c}}
\toprule
\sntd Method&Parameter&Varied?&Bounds&Value\\
\hline
--&$z$&No&--&\snz\\
--&$x_1$&No&--&\snstretch\\
\hline
Color&$t_{pk}$ (MJD)&Yes&[-15,15]+\snpeak&$59808.24^{+4.30}_{-5.59}$\\
Color&$c$&Yes&[-0.3,0.3]&$0.03^{+0.14}_{-0.06}$\\
Series&$x_0$&Yes&[0,1]&$m_B=19.62^{+0.02}_{-0.03}$\\
\hline
    \end{tabular*}

\end{table*}

\begin{figure*}
    \centering
    \includegraphics[width=\linewidth]{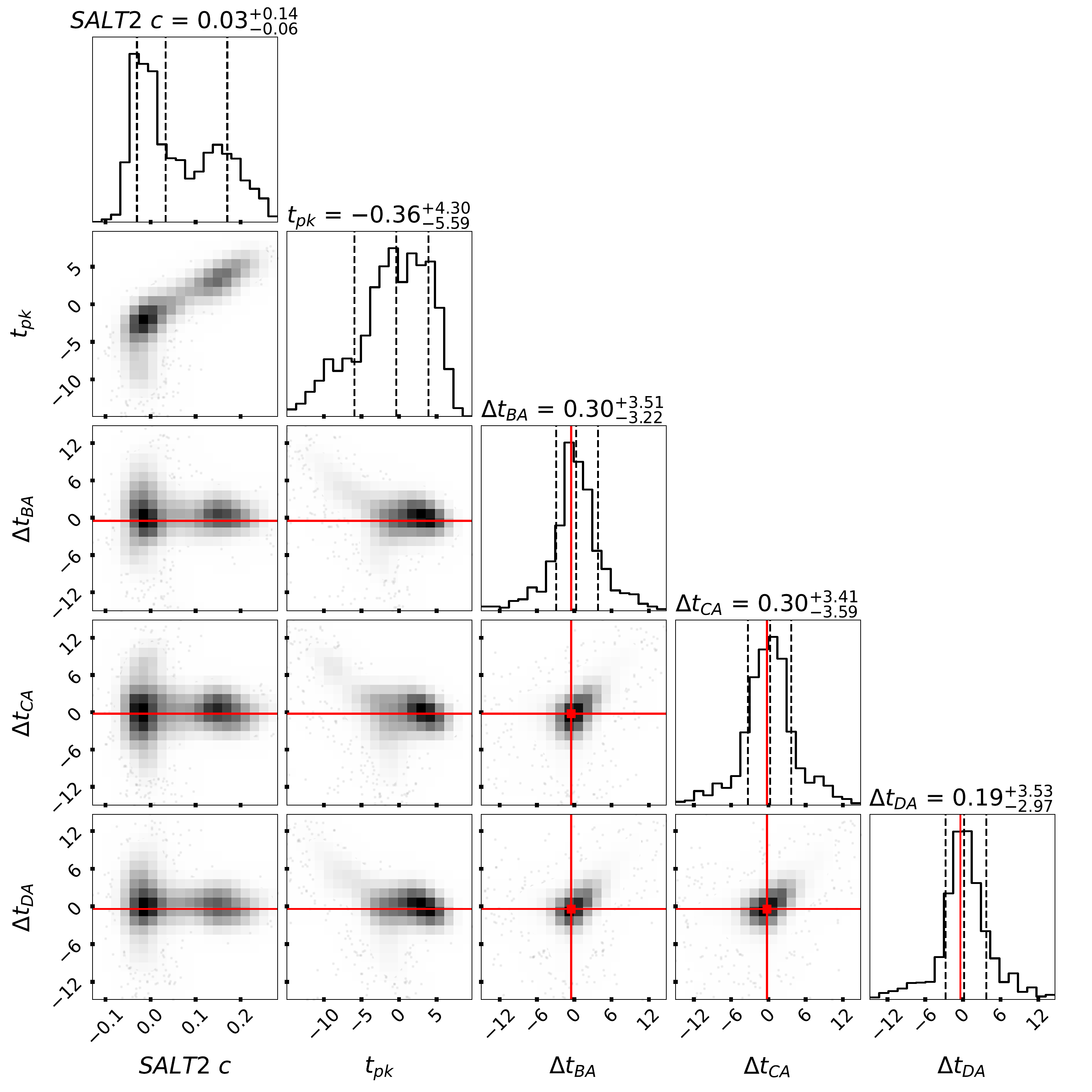}
    \caption{Posterior distributions of the \sntd Color method fitting of \hst photometry. The dashed vertical lines correspond to the distribution $16^{\rm{th}} , \ 50^{\rm{th}}, \ \rm{and} \ 84^{\rm{th}}$ quantiles and the solid red lines show the final lens model predicted time delays and magnifications (those of A and C are off of the plot, see Section \ref{sub:lens_model}). The $t_{pk}$ parameter is given relative to the G22 value of \snpeak.}
    \label{fig:sntd_corner}
\end{figure*}

\begin{figure*}
    \centering
    \includegraphics[width=\linewidth]{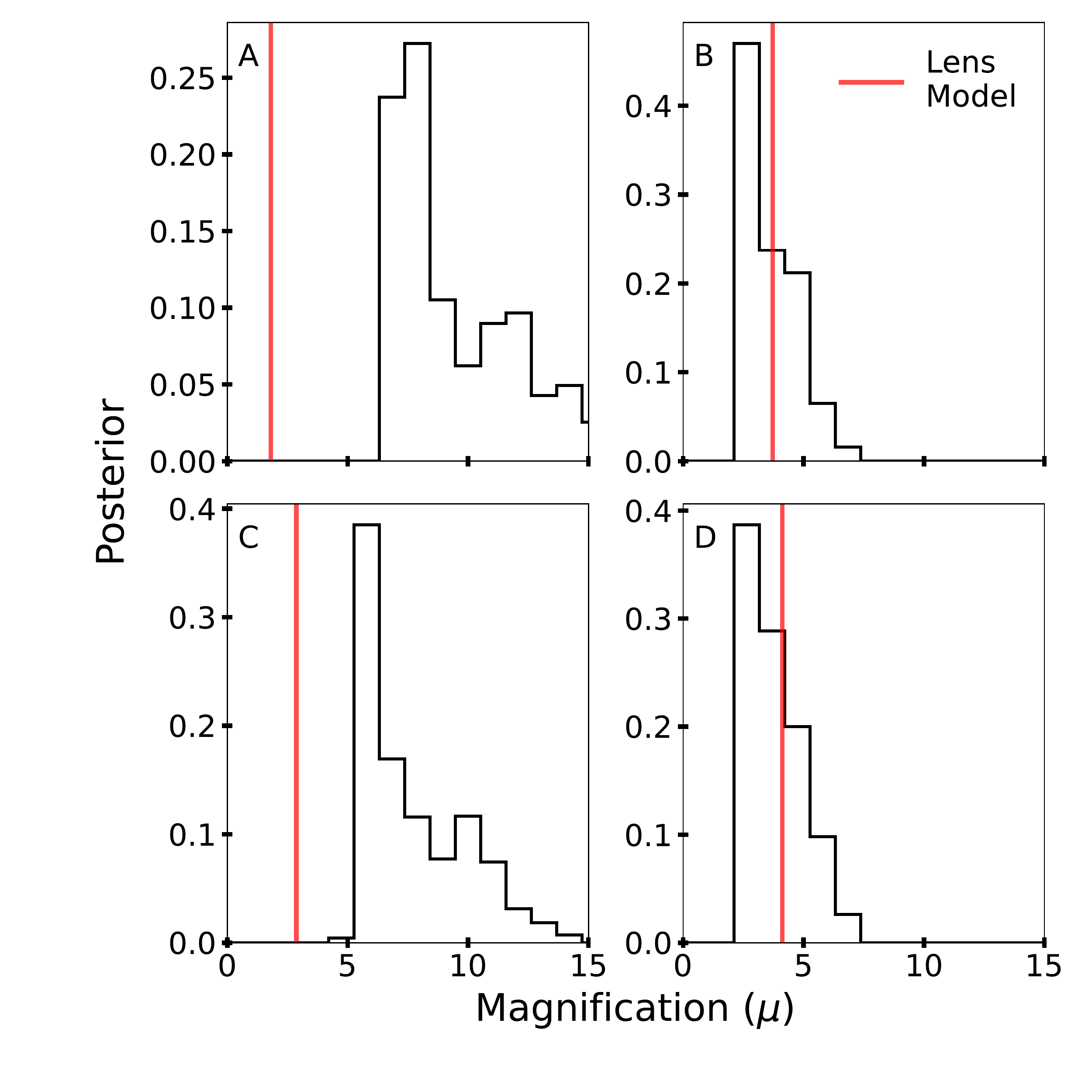}
    \caption{Posterior distributions of the magnification estimations (black), after fixing light curve and time delay parameters from Figure \ref{fig:sntd_corner}. The solid red lines show the final lens model predicted time delays and magnifications.}
    \label{fig:magnification_posteriors}
\end{figure*}

As mentioned above, \sntd measures an overall normalization ($x_0$) and relative magnifications, and so we convert the combination of $x_0$ and magnification ratios to absolute magnitudes by assuming \lensedsn is a perfect standardizable candle. Specifically, we apply light curve corrections based on Table \ref{tab:fitting} for stretch ($x_1=\snstretch$, with luminosity coefficient $\alpha=0.14$) and color ($c=0.03$ with a luminosity coefficient of $\beta=3.1$) in the manner of \citet{scolnic_complete_2018} to obtain absolute magnitude estimates. We then compare the distance modulus of each image to the value predicted by a flat $\Lambda$CDM model (with $H_0=70\,\rm{km} s^{-1} Mpc^{-1}, \ \Omega_m=0.3$) for an average SN\,Ia \citep[$M_B=-19.36$,][]{richardson_absolute_2014} at $z=\snz$, which results in a measure of the absolute magnifications. We combine the statistical uncertainties on each measured magnification with a systematic uncertainty based on the intrinsic scatter of SN\,Ia absolute magnitudes \citep[0.1\,mag;][]{scolnic_complete_2018}. The measured time delays and magnifications (with subscript ``meas'') are shown in Table \ref{tab:td_mu} compared with lens model-predicted values from Section \ref{sub:lens_results} (with subscript ``pred''). The posterior distributions for all parameters fit with \sntd (using the conversions listed above) are shown in Figures \ref{fig:sntd_corner} and \ref{fig:magnification_posteriors}. While the relative time delay uncertainties are too large to provide a useful direct cosmological constraint, these results are a valuable check on our lens modeling predictions. The agreement also supports the plausibility of measuring time delays in a single epoch, at least when the lensed SN is of Type Ia and there is some constraint on the overall explosion date.

\begin{table*}
    \centering
    \caption{\label{tab:td_mu}Measured time delays and magnifications compared to the predictions from lens models from Section \ref{sec:lens_model}.}
    
    \begin{tabular*}{\linewidth}{@{\extracolsep{\stretch{1}}}*{5}{c}}
\toprule
Image&$(\Delta t_{iA})_{\rm{meas}}$&$(\Delta t_{iA})_{\rm{pred}}$&$|\mu_{\rm{meas}}|$&$|\mu_{\rm{pred}}|$\\
&Days&Days&&\\
\hline
A&--&--&$8.31^{+4.16}_{-1.43}$
&$1.81^{+0.90}_{-0.89}$\\
B&$0.30^{+3.51}_{-3.22}$&$-0.50^{+0.15}_{-0.21}$&$3.24^{+1.69}_{-0.57}$
&$3.72^{+1.04}_{-1.24}$\\
C&$0.30^{+3.40}_{-3.59}$&$-0.22^{+0.10}_{-0.10}$&$6.73^{+3.38}_{-1.16}$
&$2.87^{+1.51}_{-1.50}$\\
D&$0.19^{+3.53}_{-2.97}$&$-0.42^{+0.12}_{-0.18}$&$3.39^{+1.65}_{-0.62}$
&$4.12^{+1.19}_{-1.36}$\\
\hline
    \end{tabular*}

\end{table*}

\section{Discussion}
\label{sec:conclusion}
We have presented the first analysis of the LensWatch collaboration, which includes the only space-based observations of the gravitationally lensed and quadruply-imaged \lensedsn. The images are resolved with \hst WFC3/UVIS (PSF FWHM $\sim0.07\arcsec$) but not with WFC3/IR (PSF FWHM $\sim0.15\arcsec$). We have measured photometry for each SN image in three optical \textit{HST} filters, and we use the resulting colors (with an additional uncertainty due to chromatic microlensing) to infer time delays ($0.30^{+3.51}_{-3.22}$, $0.30^{+3.40}_{-3.59}$, $0.19^{+3.53}_{-2.97}$ days for images B-D  relative to image A) and fluxes to infer magnification ratios using the \sntd software package. Leveraging the fact that \lensedsn is of Type Ia and therefore has a standardizable light curve, we apply a fiducial light curve standardization and obtain absolute magnification estimates of $8.31^{+4.16}_{-1.43}$, $3.24^{+1.69}_{-0.57}$, $6.73^{+3.38}_{-1.16}$, and $3.39^{+1.65}_{-0.62}$ for images A-D, respectively.

We have also carried out an analysis of the lensing system using five distinct methodologies, of which we combine four primary methods to obtain our best constraints on the magnifications, time delays, and Einstein radius ($\theta_E$) for the lensing system. We infer the smallest Einstein radius yet seen in a lensed SN ($\theta_E=(0.168^{+0.009}_{-0.005})\arcsec$) and short time delays of $\lesssim1$ day, consistent with our color curve fitting results and G22. For SN images B and D, we find consistent magnification predictions across all of our lens models that are in good agreement with the measured values. However, our lens models are unable to fully explain the observed fluxes for images A and C, and we see a significant discrepancy between measured and predicted magnification ($\sim1.7^{+0.8}_{-0.6}$\,mag and $\sim0.9^{+0.8}_{-0.6}$\,mag, respectively). 

We resort to variations from microlensing and/or millilensing \citep{metcalf_flux_2002,foxley-marrable_impact_2018,goldstein_precise_2018,hsueh_flux-ratio_2018,huber_strongly_2019} to explain this inconsistency. We do see some evidence for differential dust extinction and/or chromatic microlensing across the four images of \lensedsn with relative differences in measured F475W-F814W of up to $\sim0.18\pm0.08$\,mag, and based on the work of \citet{goldstein_precise_2018} our \hst epoch is $\sim3$ rest-frame weeks outside of the ``achromatic phase'' of SN\,Ia microlensing where the impact on optical colors is expected to be $\lesssim0.1$\,mag (95\% confidence). We include the additional predicted uncertainty due to microlensing in our fitting. However, the F475W filter was most discrepant with the SALT2 fitting suggesting a possible systematic error in the photometry, and regardless this differential extinction is insufficient to explain the discrepancies we observe in images A and C relative to B (the largest color difference). We therefore estimate the additional (roughly achromatic) magnification to be $\sim1$ magnitude in both images, which is significant but well within expectations for average galaxy-scale microlensing \citep{pierel_projected_2021} or millilensing \citep{metcalf_flux_2002}. While it is suggested that saddle images are more likely to be demagnified by microlensing \citep[e.g.,][]{schechter_quasar_2002}, we expect the method of detection for \lensedsn would be significantly biased toward microlensing events with high magnifications. A template epoch is already scheduled for after \lensedsn will have faded, which will drastically simplify and improve the photometry and lens modeling processes and provide more stringent constraints.

Although the multiply-imaged SN population is still small, this decade it is expected to grow by orders of magnitude with the Vera C. Rubin Observatory Legacy Survey of Space and Time \citep[LSST;][]{ivezic_lsst_2019} and \textit{Nancy Grace Roman Space Telescope} \citep{pierel_projected_2021}. Though both telescopes are expected to find a large number of spatially resolved lensing systems, unresolved lensed SNe discovered with Rubin will still be common, and a dedicated follow-up campaign from space similar to LensWatch will be necessary to provide accurate photometry and lens modeling for such systems. \lensedsn is the first lensed SN analysis presented by the LensWatch program, and is an excellent example of the coordination that will be required for upcoming lensed SN cosmology efforts. The unresolved, ground-based discovery with ZTF and subsequent follow-up with \textit{HST} is a glimpse at likely future discoveries with Rubin and follow-up with \textit{Roman} (or \textit{HST}), a strategy that can be extremely fruitful for the field of gravitationally lensed SN cosmology.

\clearpage

\begin{center}
    \textbf{Acknowledgements}
\end{center}
We would like to thank David Jones, Nao Suzuki and Taylor Hoyt for discussions helpful to the analysis in this work. 
This paper is based in part on observations with the NASA/ESA Hubble Space Telescope obtained from the Mikulski Archive for Space Telescopes at STScI. The specific observations analyzed can be accessed via \dataset[DOI]{https://doi.org/10.17909/gvys-q017}; support was provided to JDRP and ME through program HST-GO-16264. SE and SHS thank the Max Planck Society for support through the Max
Planck Research Group for SHS. Support for this work was provided by NASA through the NASA Hubble Fellowship grant HST-HF2-51492 awarded to AJS by the Space Telescope Science Institute, which is operated by the Association of Universities for Research in Astronomy, Inc., for NASA, under contract NAS5-26555. This work has also been enabled by support from the research project grant ‘Understanding the Dynamic Universe’ funded by the Knut and Alice Wallenberg Foundation under Dnr KAW 2018.0067.  This project has received funding from the European Research Council (ERC)
under the European Union’s Horizon 2020 research and innovation programme (LENSNOVA: grant agreement No 771776).
This research is supported in part by the Excellence Cluster ORIGINS which is funded by the Deutsche Forschungsgemeinschaft (DFG, German Research Foundation) under Germany's Excellence Strategy -- EXC-2094 -- 390783311. SS acknowledges financial support through grants PRIN-MIUR 2017WSCC32 and 2020SKSTHZ. Part of this research was carried out at the Jet Propulsion Laboratory, California Institute of Technology, under a contract with the National Aeronautics and Space Administration (80NM0018D0004). JH, CC and RW were supported by a VILLUM FONDEN Investigator grant to JH (project number 16599). XH was supported in part by the University of San Francisco Faculty Development Fund. GL and YS acknowledge the support from the China Manned Spaced Project (CMS-CSST-2021-A12).

FP acknowledges support from the Spanish State Research Agency (AEI) under grant number PID2019-105552RB-C43. SD acknowledges support from the European Union's Horizon 2020 research and innovation programme Marie Skłodowska-Curie Individual Fellowship (grant agreement No. 890695), and a Junior Research Fellowship at Lucy Cavendish College, Cambridge. TP acknowledges the financial support from the Slovenian Research Agency (grants I0-0033, P1-0031, J1-8136 and Z1-1853). This work was supported by a collaborative visit funded by the Slovenian Research Agency (ARRS, travel grant number BI-US/22-24-006). IP-F acknowledges support from the Spanish State Research Agency (AEI) under grant number PID2019-105552RB-C43. CG is supported by a VILLUM FONDEN Young Investigator Grant (project number 25501). CL acknowledges support from the National Science Foundation Graduate Research Fellowship under Grant No. DGE-2233066.


\clearpage

\bibliographystyle{aasjournal}

\appendix

\textbf{Individual Lens Modeling Results:}\\

\section{Modeling with \GLEE}
 
\subsection{Lens model parameterization}
We modeled SN Zwicky with an automated modeling pipeline that is based on the modeling software \GLEE \citep{ertl_tdcosmo_2022, suyu_halos_2010, suyu_disentangling_2012}, where we adopt the SPEMD \citep{barkana_fast_1998} profile whose dimensionless surface mass density (or convergence) is given by 
\begin{equation}
\label{eq:kappa_spemd}
	\kappa_{\rm SPEMD}(x, y) = \frac{3-\gamma}{2}\left[\frac{\theta_{\rm E}}{\sqrt{q_{\rm m}(x-x_{\rm m})^2+\frac{(y-y_{\rm m})^2}{q_{\rm m}}}}\right]^{\gamma-1},
\end{equation}
where $(x_{\rm m}, y_{\rm m})$ is the lens mass centroid, $q_{\rm m}$ is the axis ratio of the elliptical mass distribution, $\theta_{\rm E}$ the Einstein radius, and $\gamma$ is the power-law slope.  The mass distribution is then rotated by the position angle $\phi_{\rm m}$, where an elliptical mass distribution with an angle of $\phi_{\rm m} = 0$ corresponds to elongation along the $y$-axis (after converting to the conventional definition of position angle). 
The external shear strength is described by $\gamma_{\rm ext} = \sqrt{\gamma_{\rm ext,1}^2 + \gamma_{\rm ext,2}^2}$, with $\gamma_{\rm ext,1}$ and $\gamma_{\rm ext,2}$ the components of the shear. For a shear position angle $\phi_{\rm ext}=0$, the system is sheared along the $x$-direction.

\subsection{Results based on SN image positions}
\begin{figure}
    \centering
    \includegraphics[width=\columnwidth]{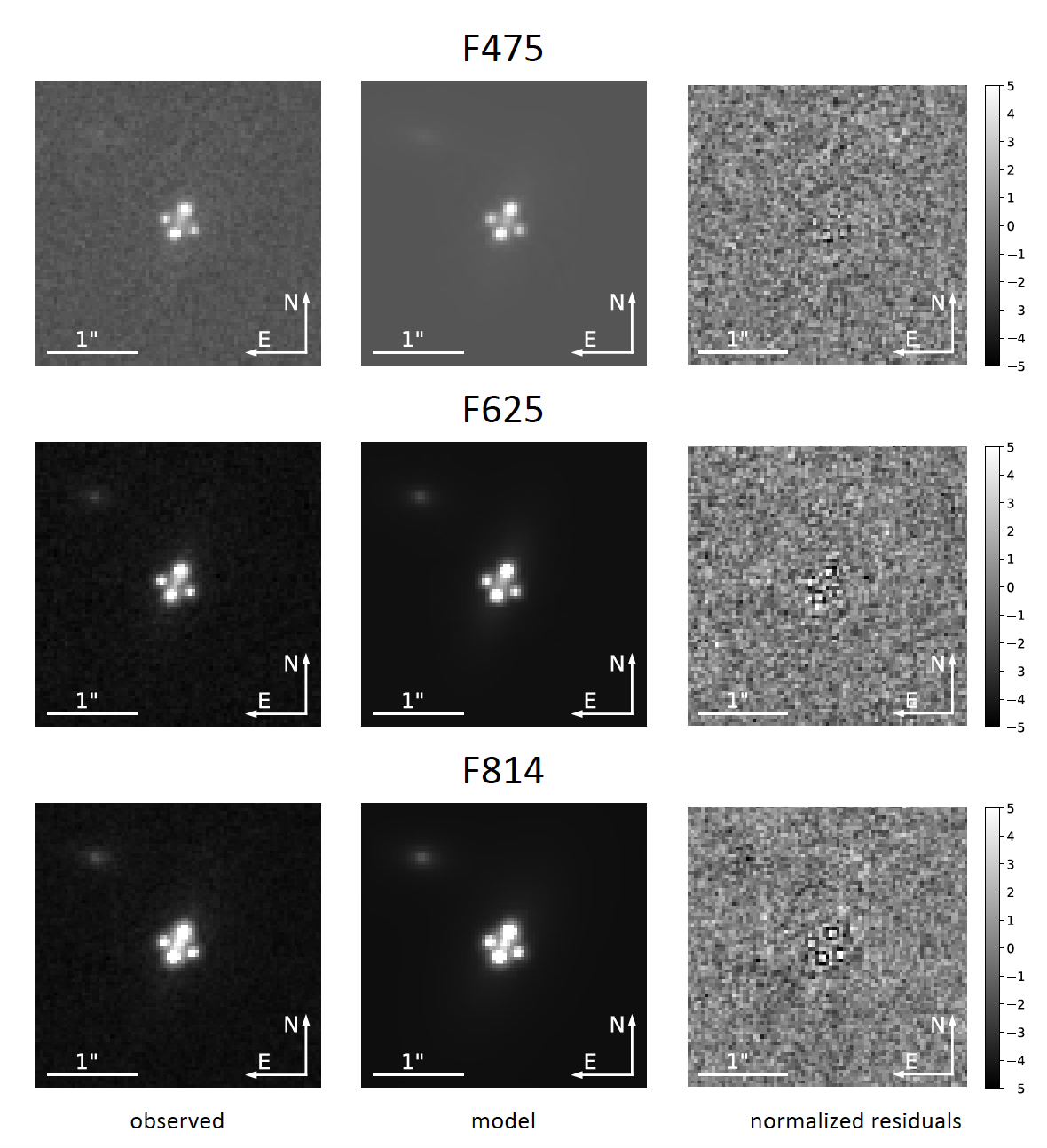}
    \caption{\label{fig:glee_sbfit} \GLEE: Surface brightness fitting with \GLEE\ in the three HST filters, as shown in the different rows. From left to right: observed image, model, and normalized residuals after modeling the light of the lens galaxy and the four SN images.}
\end{figure}

First, we model the light of the lens galaxy with two S\'{e}rsic profiles, and the light of multiple lensed SN images by fitting a PSF model constructed from multiple stars in the field of the drizzled data. \citet{ertl_tdcosmo_2022} showed that for lensed quasars we can achieve astrometric accuracy of 2 milli-arcseconds (mas) from the surface brightness (SB) fit, by comparing the modeled image positions to those measured by the Gaia satellite. We use our SN image positions (from PSF fitting) to constrain the mass parameters, since we did not find (and do not expect) any substantial lensed arc light (from the SN host galaxy) in the modeling residuals of the three UVIS bands. We show the results of our SB fit in Fig.~\ref{fig:glee_sbfit}.
The measured astrometric positions of the four SN images in all three modeled bands are summarized in Tab.~\ref{tab:appendix_glee_astrometry}, and the lens light properties (based on the first and second brightness moments of the modeled lens light distribution that is a combination of the 2 S\'{e}rsics) in the F814W filter is in Tab.~\ref{tab:appendix_glee_sersic_results}. The positions in the F475W and F625W bands are aligned with the F814W coordinate frame.

\begin{table*}[h]
    \centering
    \caption{\label{tab:appendix_glee_astrometry}\GLEE: Astrometry and brightness of SN images -- best-fit SN image positions and amplitudes from surface brightness fitting.}
    
    \begin{tabular*}{\linewidth}{@{\extracolsep{\stretch{1}}}*{5}{c}}
\toprule
Image&&F475&F625&F814\\
\hline
&x[\arcsec]&1.688&1.691&1.686\\
A&y[\arcsec]&1.781  &1.783& 1.783\\
&amplitude&9.96  &29.11& 62.60\\ \\
&x[\arcsec]&1.789&1.788&1.791\\
B&y[\arcsec]&1.545& 1.547& 1.549\\
&amplitude&4.17  &12.36& 23.61\\ \\
&x[\arcsec]&1.575&1.577&1.576\\
C&y[\arcsec]&1.504&1.504& 1.500\\
&amplitude&9.08  &27.39& 51.27\\ \\
&x[\arcsec]&1.470 &1.466& 1.469\\
D&y[\arcsec]&1.671&1.667& 1.669\\
&amplitude&4.62  &12.72& 25.49\\
\hline
    \end{tabular*}

\end{table*}

For each band, we use the image positions reported in Tab.~\ref{tab:appendix_glee_astrometry} and adopt an uncertainty on the image positions of 4 mas to constrain the lens mass parameters. The 4 mas is an estimate based on the astrometric accuracy of 2 mas \citep{ertl_tdcosmo_2022} and to account for substructure lensing, which can perturb the image positions at the few mas level, as shown by \citet{chen_astrometric_2007}. We impose uniform priors on all 8 lens mass parameters that are tabulated in the leftmost column of Tab.~\ref{tab:appendix_glee_results}. We report the mass model results from the F814W band because we achieved the lowest image-position $\chi^{2}$ in this band. The results in this band are consistent with those of the other 2 bands, and also with the results of our light fit of the lens galaxy. We do not combine the constraints from all 3 bands in one single model, because positional uncertainties due to astrometric perturbations from e.g.~substructures in the mass distributions are dominant and the measured SN image positions of the individual bands are thus not completely independent.
Our final lens mass and shear parameters are presented in Tab.~\ref{tab:appendix_glee_results}.
Comparing to the modeled lens light in Tab.~\ref{tab:appendix_glee_sersic_results}, the lens galaxy mass profile agrees well with the light in terms of centroid, axis ratio and position angle.
In Tab.~\ref{tab:appendix_glee_kappa}, we present the convergence $\kappa$ and total shear strength $\gamma_{\rm{tot}}$ at the (modeled) image positions.

\begin{table*}[h]
    \centering
    \caption{\label{tab:appendix_glee_sersic_results}\GLEE: Centroid, axis-ratio, and position angle of the lens light computed from the second brightness moments of the two Sersic profiles in our best-fit model of the F814W filter.}
    
    \begin{tabular*}{.32\linewidth}{llc}
\toprule
Parameter&Description&\\
\hline
$x_{\rm{S}}$[\arcsec]&$x$-centroid&1.641\\
$y_{\rm{S}}$[\arcsec]&$y$-centroid&1.650\\
$q_{\rm{S}}$&axis-ratio&0.52\\
$\phi_{\rm{S}}$ [deg]&position angle&155\\
\hline
    \end{tabular*}

\end{table*}

\begin{table*}[h]
    \centering
    \caption{\label{tab:appendix_glee_results}\GLEE: Modeled mass and shear parameters in the F814W band. We present the median and 1$\sigma$ uncertainties. Position angles are reported as East of North.}
    
    \begin{tabular*}{.42\linewidth}{llc}
\toprule
Parameter&Description&\\
\hline
$x_{\rm{m}}$[\arcsec]&$x$-centroid&$1.645^{+0.002}_{-0.002}$\\
$y_{\rm {m}}$[\arcsec]&$y$-centroid&$1.657^{+0.006}_{-0.004}$\\
$q_{\rm {m}}$&axis ratio&$0.47^{+0.18}_{-0.13}$\\
$\phi_{\rm {m}}$ [deg]&position angle&$159^{+2}_{-2}$\\
$\theta_{\rm{E}}$[\arcsec]&Einstein radius&$0.168^{+0.005}_{-0.004}$\\
$\gamma$&power-law index&$2.04^{+0.14}_{-0.22}$\\
\hline
$\gamma_{\rm{ext}}$&shear strength&$0.02^{+0.03}_{-0.01}$\\
$\phi_{\rm{ext}}$ [deg]&shear position angle&$52^{+37}_{-57}$\\
\hline
    \end{tabular*}
\end{table*}

\begin{table*}[h]
    \centering
    \caption{\label{tab:appendix_glee_kappa}\GLEE: Convergence $\kappa$ and total shear strength $\gamma_{\rm tot}$ at the (modeled) image positions.}
    
    \begin{tabular*}{0.6\linewidth}{@{\extracolsep{\stretch{1}}}*{3}{c}}
\toprule
Image&$\kappa$ & $\gamma_{\rm tot}$\\
\hline
A&$0.75^{+0.15}_{-0.15}$& $0.95^{+0.26}_{-0.24}$\\
B&$0.26^{+0.13}_{-0.08}$& $0.43^{+0.04}_{-0.06}$\\
C&$0.65^{+0.09}_{-0.12}$& $0.81^{+0.18}_{-0.18}$\\
D&$0.27^{+0.13}_{-0.08}$& $0.45^{+0.05}_{-0.07}$\\
\hline
    \end{tabular*}

\end{table*}

\subsection{Impact on results due to flux constraints}
To investigate the bias to higher magnification for images A and C, we include fluxes, which were obtained from the light fit to the SN images (listed in Tab.~\ref{tab:appendix_glee_astrometry}), in our model. The flux amplitude had typical uncertainties of $\sim$$1\%$. 
We find that models based on image positions and fluxes try to fit to the fluxes at the expense of the poorer image position recovery, so the model cannot fit well to both positions and fluxes.

The image positions are close to a critical curve, so small shifts lead to large change in magnification. This is especially evident for the model where we use flux uncertainties from the SB fit. Imposing higher flux uncertainties (either 10\% or 20\% of the modeled flux value) leads to a lower image position $\chi^{2}$ and a higher magnification $\chi^{2}$ and brings the modeling results closer to the models where we used only image positions.

\begin{figure}
    \centering
    \includegraphics[trim={0 1cm 0 2cm},clip,width=0.8\columnwidth]{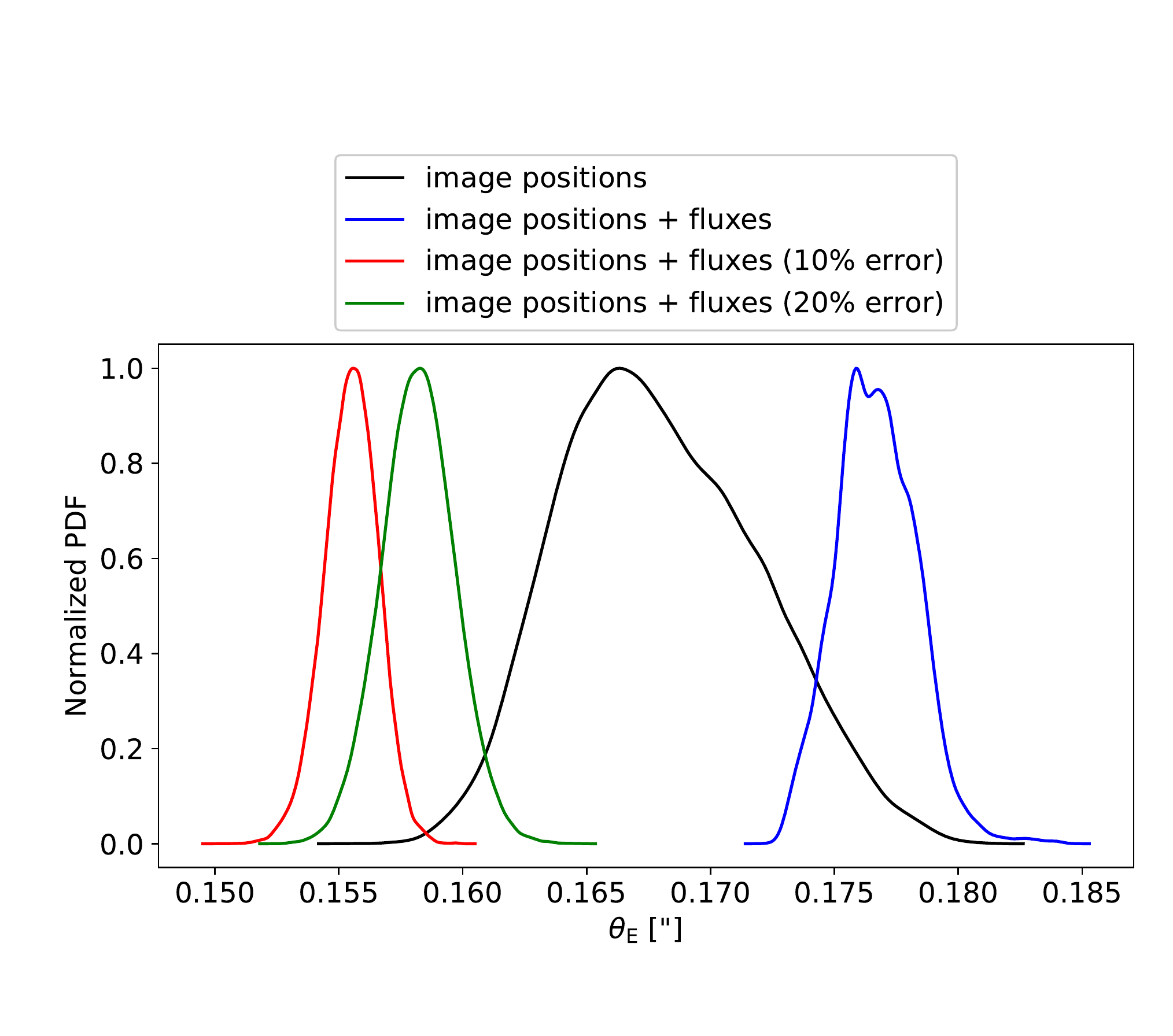}
    \caption{\label{fig:appendix_glee_fluxes1}Exploration of the impact of flux constraints on the lens models with \GLEE: Distribution of Einstein radii for the 4 different model classes. The flux constraints are based on the measured SN image amplitudes in Tab.~\ref{tab:appendix_glee_astrometry}, with typical uncertainties of $\sim$1\%, unless boosted to 10\% or 20\% as indicated in the legend.  }
\end{figure}

\begin{figure}
    \centering
    \includegraphics[width=\columnwidth]{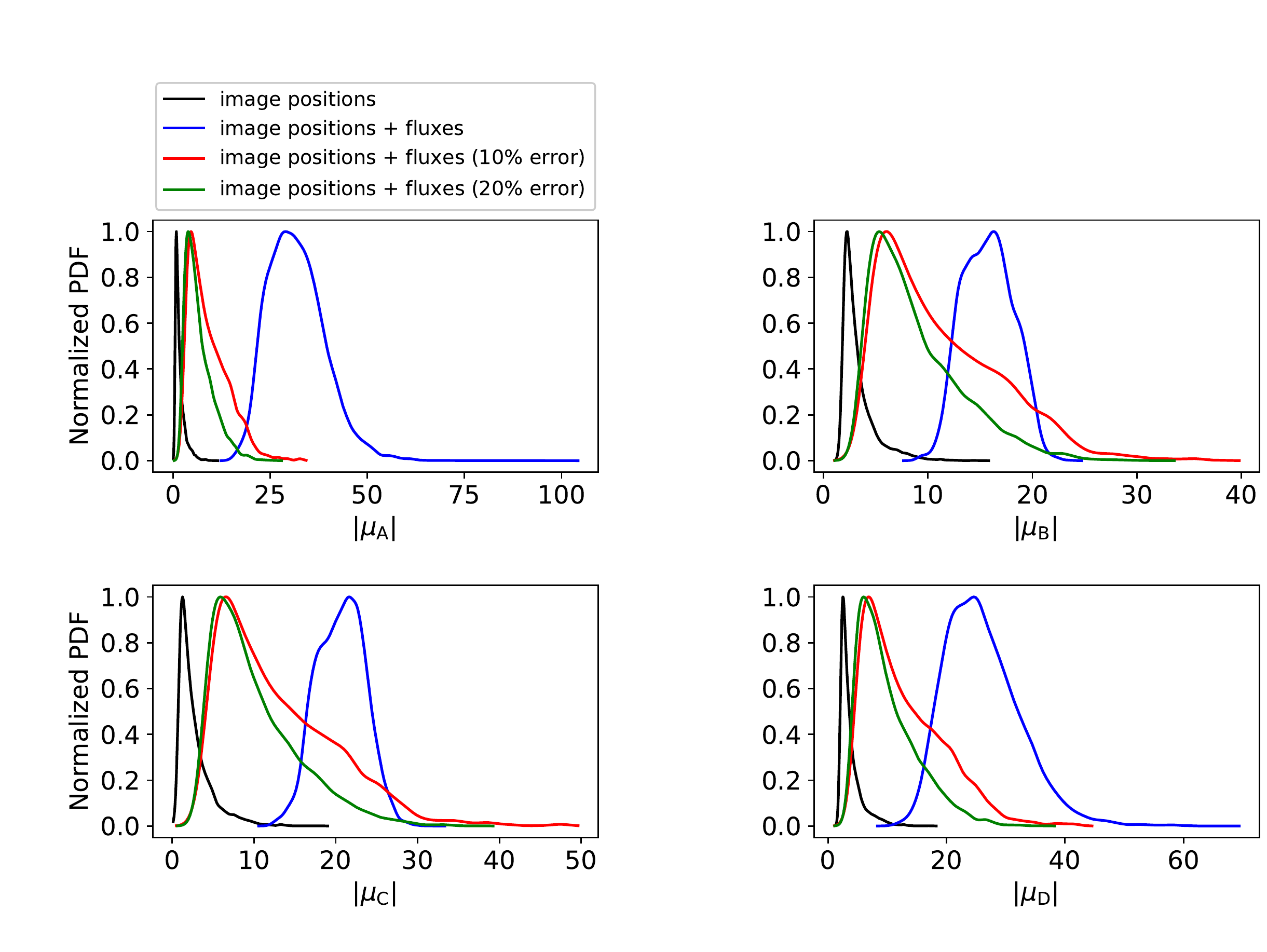}
    \caption{\label{fig:appendix_glee_fluxes2}Similar to Fig.~\ref{fig:appendix_glee_fluxes1} from \GLEE but for the SN image  magnifications from the 4 different model classes.}
\end{figure}

\begin{figure}
    \centering
    \includegraphics[trim={2cm 0 1cm 0},clip,width=\columnwidth]{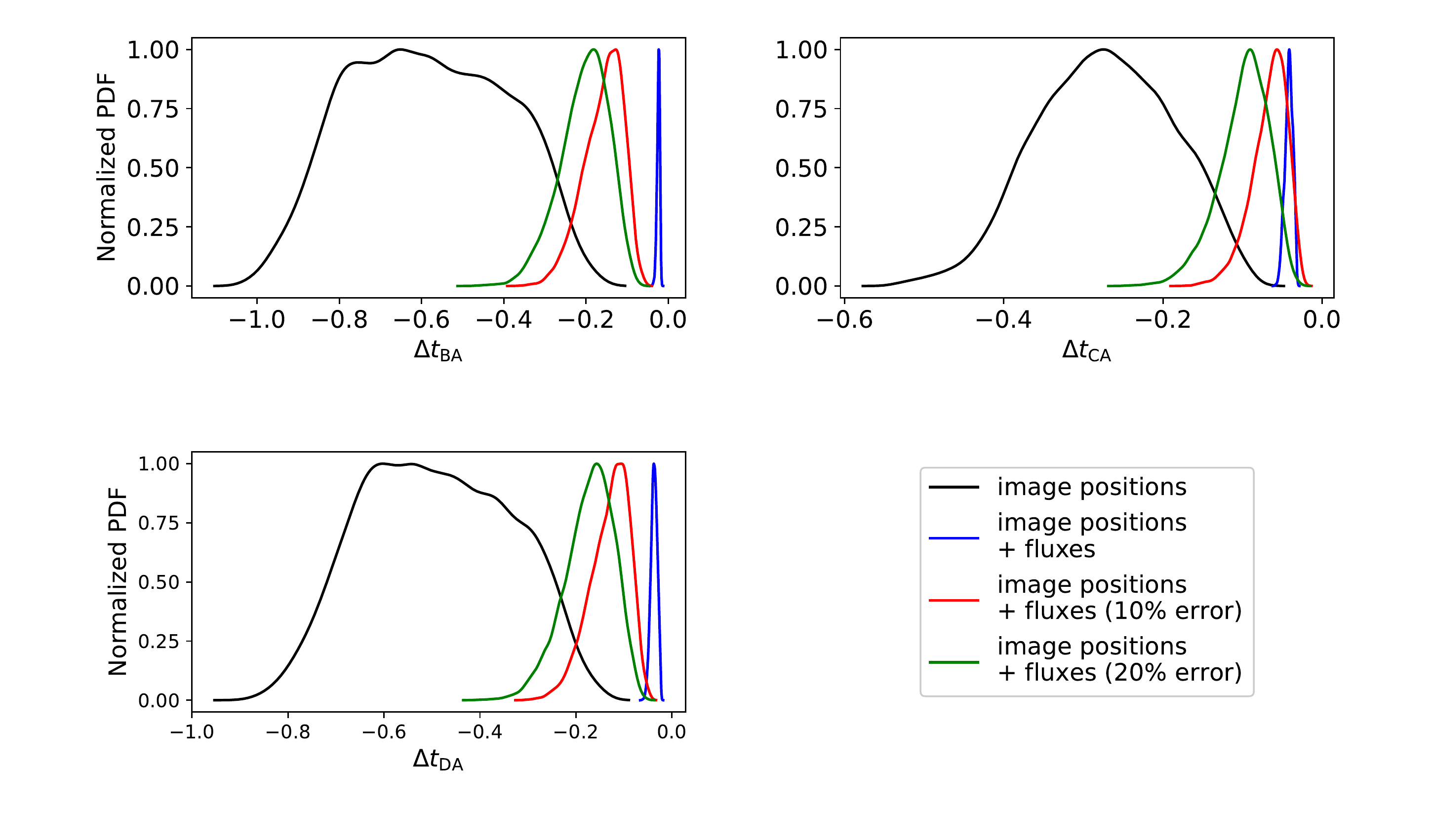}
    \caption{\label{fig:appendix_glee_fluxes3}Similar to Fig.~\ref{fig:appendix_glee_fluxes1} from \GLEE but for the distribution of predicted time delays from the 4 different model classes.}
\end{figure}

We show the impact of including fluxes in our model by plotting the distribution of $\theta_{\rm E}$ in Fig.~\ref{fig:appendix_glee_fluxes1}, magnifications in Fig.~\ref{fig:appendix_glee_fluxes2}, and predicted time delays in Fig.~\ref{fig:appendix_glee_fluxes3} for the 4 different model classes. Since the models with flux constraints do not fit well to both the image positions and fluxes (with $\chi^2 \gtrsim 10$), these models result in underestimated mass parameter uncertainties, as indicated by the narrower distributions for the blue, red and green models in Fig.~\ref{fig:appendix_glee_fluxes1}.   


\section{Modelling with {\tt lfit\_gui}}

\begin{figure}[htbp]
\centering
\includegraphics[width=0.32\textwidth]{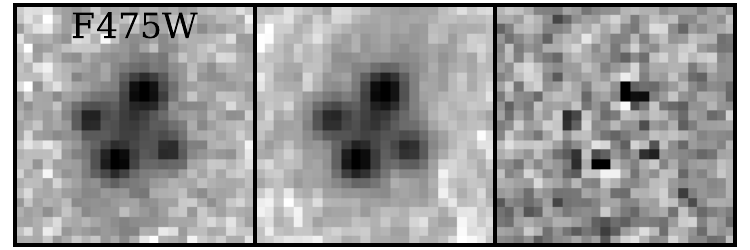}
\includegraphics[width=0.32\textwidth]{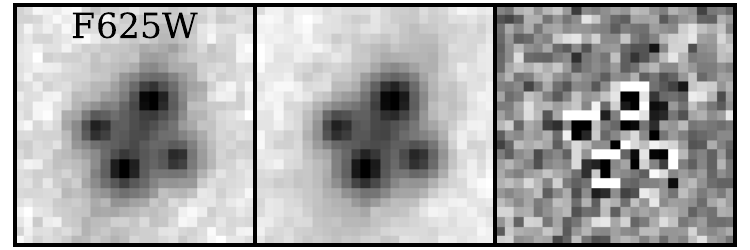}
\includegraphics[width=0.32\textwidth]{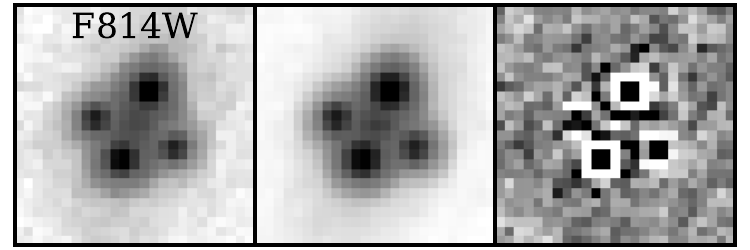}
\caption{Photometric fitting results in the F475W (left), F625W (middle), and F814W (right) in the {\tt lfit\_gui} approach (every cutout is approximately 1 arcsec by 1 arcsec). In each sub-panel, data, best-fit model, and residual are shown from left to right.}
\label{fig:photometric_fit_YS}
\end{figure}

{\tt lfit\_gui} is a lens modeling software introduced by \citet{shu_lfit_2016}, which has been applied to about 150 strong-lens systems \citep{shu_lfit_2016,shu_boss_2016,shu_sloan_2017,marques-chaves_discovery_2017,wang_sdss_2017,shu_prediction_2018,marques-chaves_rest-frame_2020}. In order to maintain an independent analysis, in the {\tt lfit\_gui} approach, positions and fluxes of the four SN images in the three optical filter bands are independently measured by fitting a photometric model consisting of two concentric Sersic components, four PSF components, and a constant component to the drizzled data downloaded from the Barbara A. Mikulski Archive for Space Telescopes Portal\footnote{The .drc files from https://mast.stsci.edu/portal/Mashup/Clients/Mast/Portal.html}. The PSF models constructed in the \GLEE approach are used. These photometric fitting results are shown in Figure~\ref{fig:photometric_fit_YS}. Overall speaking, the photometric model considered is able to reproduce the main structures in the data. Some residuals are seen at the lensed SN positions, which are primarily caused by PSF mismatches. The measured positions and fluxes of the four SN images are summarised in Table~\ref{tb:photometric_fit_YS}. The positional uncertainties are clearly correlated with the signal-to-noise ratios (S/Ns) of the lensed SN images. As a reulst. they are the smallest in F814W ($\approx 1$mas) and the highest in F475W ($\approx 4$mas). In general, the measured positions of the four lensed SN images agree well across the three bands. The largest differences are seen in the relative $x$ positions of images C and D between F475W and F625W, which are about $1.8\sigma$. The measured photometry for the four lensed SN images are found to be systematically brighter than the measurements in Table~\ref{tab:im_mags}. The differences are typically within 0.15 mag in F625W and F814W and become 0.3--0.6 mag in F475W. We think this is likely related to the different treatments of the lens galaxy light. It affects photometry in the F475W the most because the brightness contrast between lensed SN images and the lens galaxy is the smallest.

\begin{figure}[htbp]
\centering
\includegraphics[width=0.73\textwidth]{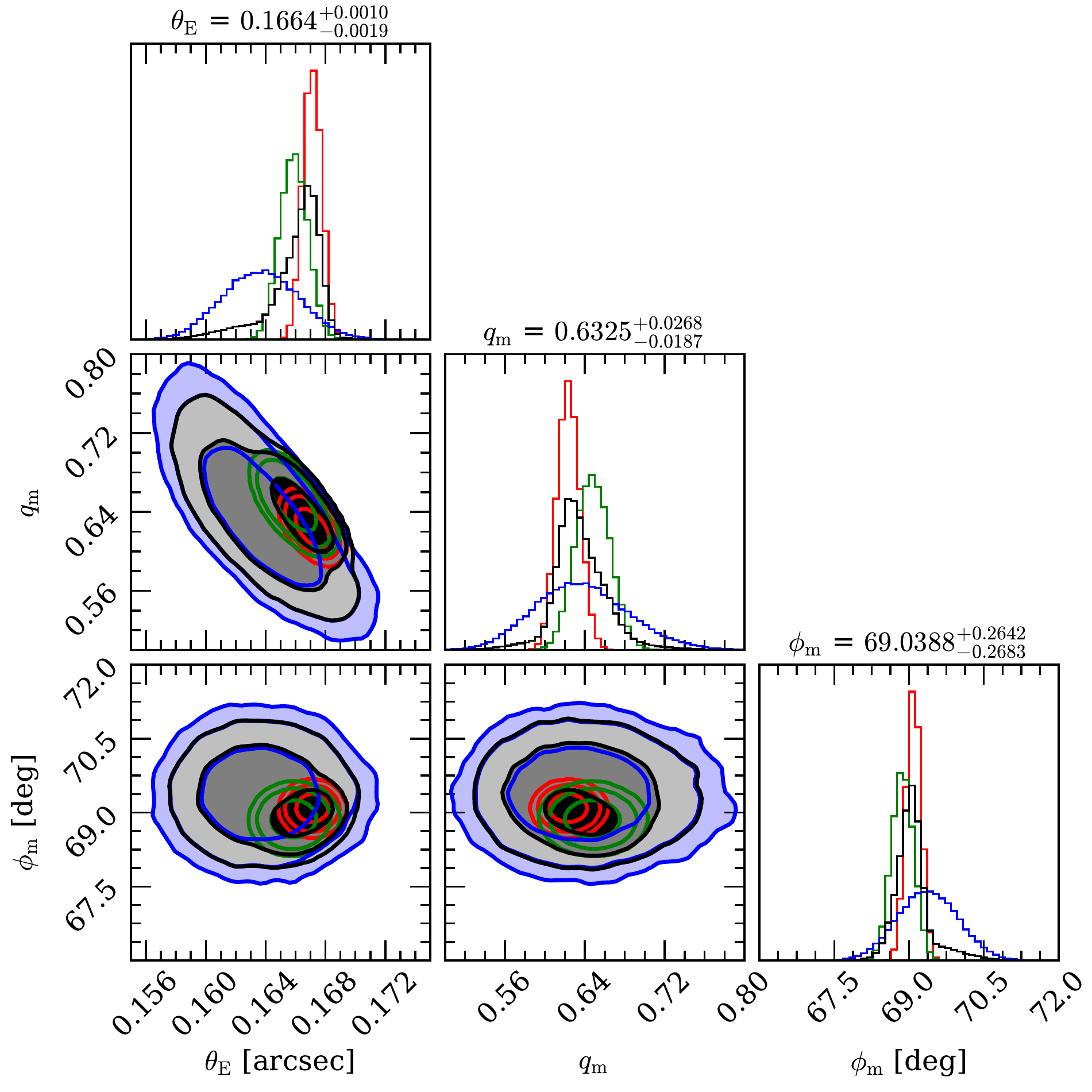}
\caption{Posterior PDFs for the key SIE model parameter in the {\tt lfit\_gui} approach. The blue, green, and red contours and histograms correspond to results inferred from the F475W, F625W, and F814W data, and the black contours and histograms correspond to the combined results.}
\label{fig:corner_plot_YS}
\end{figure}

\begin{figure}
\centering
\includegraphics[width=0.32\textwidth]{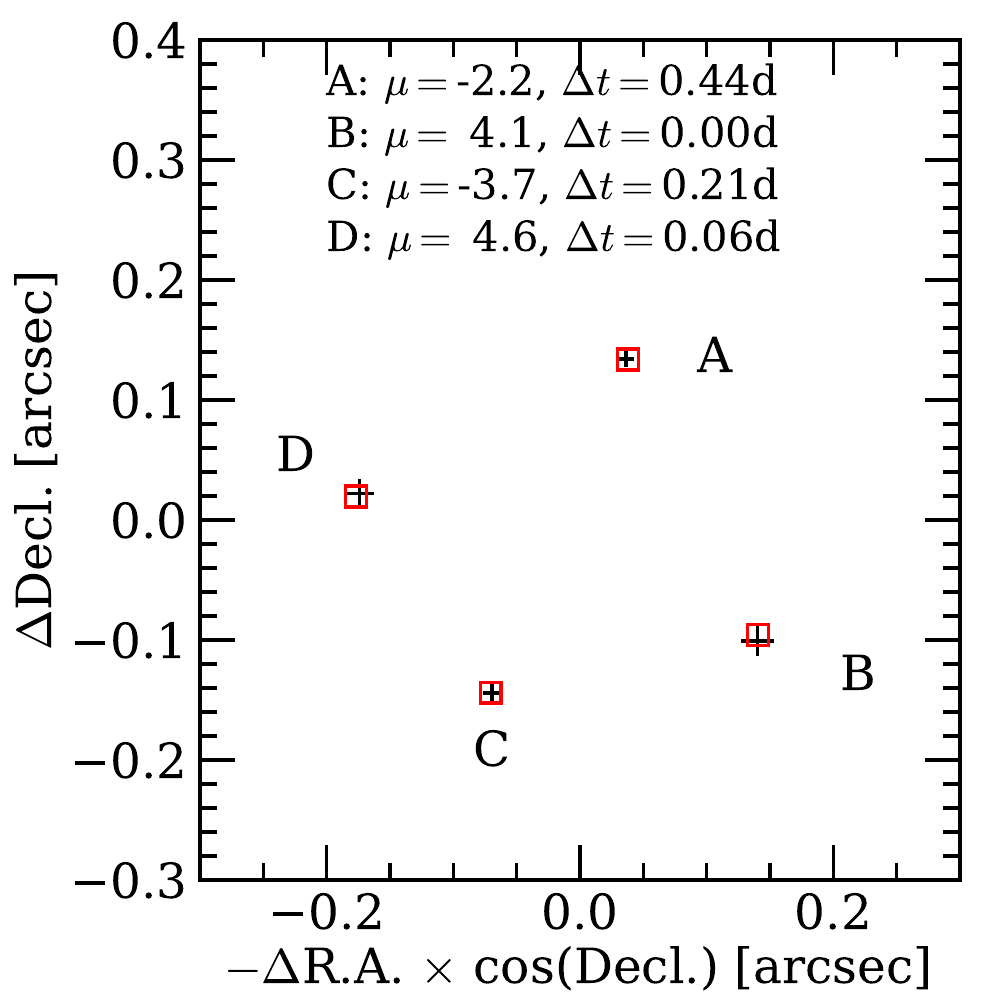}
\includegraphics[width=0.32\textwidth]{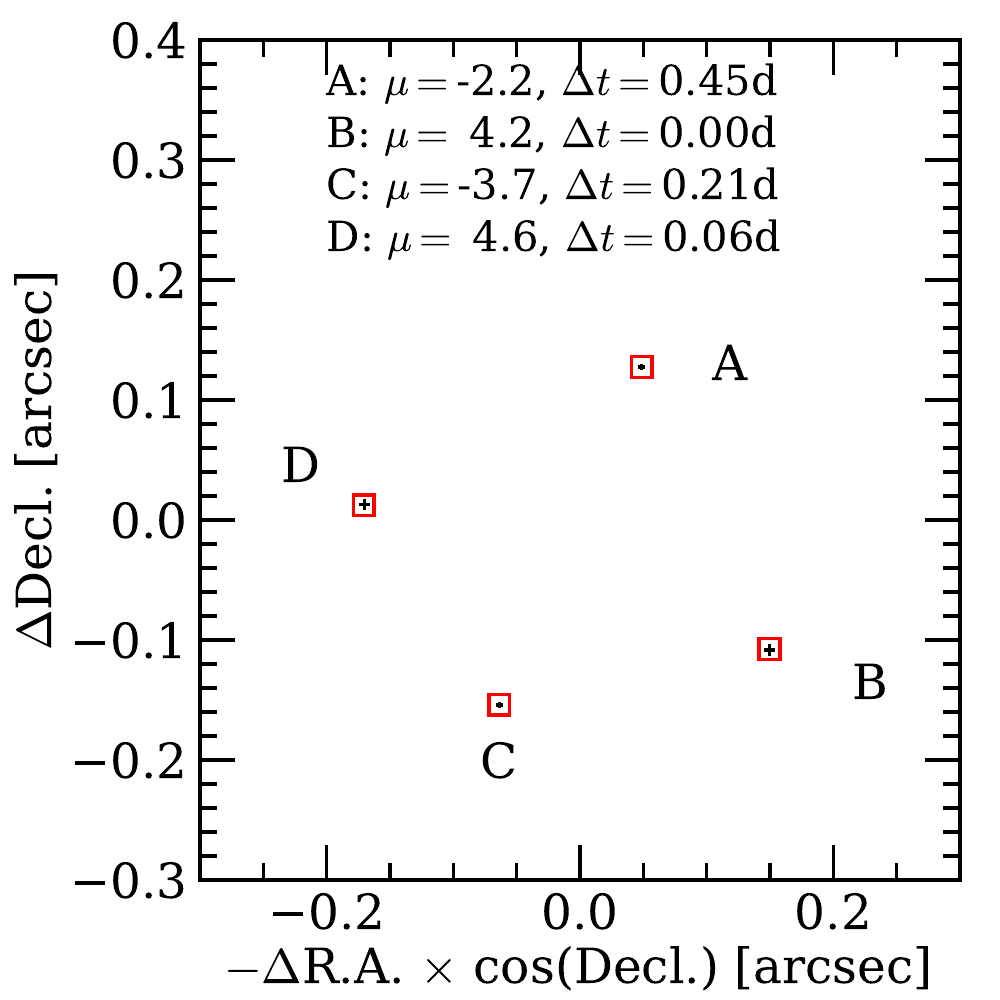}
\includegraphics[width=0.32\textwidth]{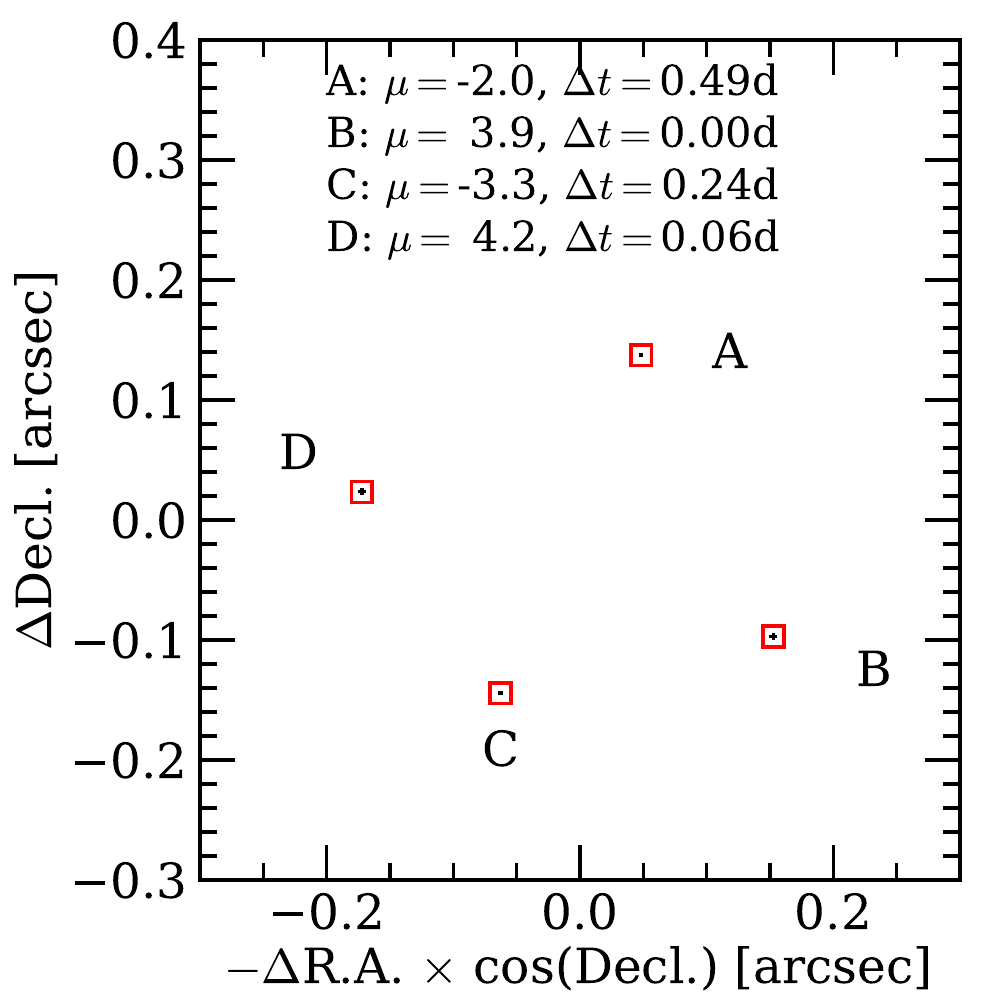}
\caption{Modeling results using the MAP solution in the F475W (left), F625W (middle), and F814W (right) in the {\tt lfit\_gui} approach. Black symbols with error bars ($3\sigma$) indicate the measured SN positions, and the red squares correspond to the predicted image positions. The predicted magnifications and relative time delays are also shown.}
\label{fig:sie_models_YS}
\end{figure}

In terms of lens modeling, the {\tt lfit\_gui} approach considered an SIE lens model, the convergence of which follows the profile defined in Equation~\ref{eq:kappa_spemd} but with $\gamma$ fixed to 2, and used the measured positions of the four SN images to constrain the five SIE parameters (as well as the source position) in the three bands separately. The sampling was done using the EnsembleSampler from the {\tt emcee} package assuming uniform priors with sufficiently wide ranges for all the seven free parameters. The maximum a posteriori (MAP) estimation and marginalised posterior distribution for the three key SIE parameters, i.e. Einstein radius, axis ratio, and position angle, are reported in Table~\ref{tab:sie_pars_YS}, and the posterior probability density distributions (PDFs) are provided in Figure~\ref{fig:corner_plot_YS}. As shown in Figure~\ref{fig:sie_models_YS}, this lens model well reproduces the four lensed SN positions. The root mean square of the differences between the predicted and observed image positions is $0.002^{\prime \prime}$, $0.0004^{\prime \prime}$, and $0.0001^{\prime \prime}$ in F475W, F625W, and F814W respectively. The tightest constraints on the lens model parameters are obtained in F814W that has the smallest positional uncertainties and the posterior PDFs are the broadest in F475W. Nevertheless, the lens model parameters are generally consistent within $1\sigma$. We find a clear anti-correlation between the Einstein radius and axis ratio (Figure~\ref{fig:corner_plot_YS}), which is also observed in other lens modelling methods (e.g. Figure~\ref{fig:LS2_posteriors} and Figure~\ref{fig:salt+ls-corner}).

We use the Bayesian information criterion (BIC) to combine results from the three bands, which is defined as 
\begin{equation}
    \text{BIC}=p\ln(n) -2\ln(\widehat{L}),
\end{equation}
where $p$ is the number of free parameters (i.e. 7), $n$ is the number of constraints (i.e. 8), and $\widehat{L}$ is the maximum likelihood of a model. The BIC values are 15.9757, 14.9495, and 14.6745 in the F475W, F625W, and F814W. Weighting the results in F475W and F625W relative to F814W as 
\begin{equation}
    \text{weight}_{\rm F475W/F625W} = \exp(-0.5 \times (\text{BIC}_{\rm F475W/F625W}-\text{BIC}_{\rm F814W})),
\end{equation}
the combined results suggest that the Einstein radius is $0.1664_{-0.0019}^{+0.0010}$ arcsec, the axis ratio is $0.632_{-0.019}^{+0.027}$, and the position angle is $69.0_{-0.3}^{+0.3}$ degrees (i.e. the angle between the major axis of the lens surface mass density distribution and the x axis, measured counterclockwise). The total lensing mass (within the ellipse that corresponds to $\kappa =1$) is thus estimated to be $7.47_{-0.17}^{+0.09} \times 10^{9} M_\odot$. The predicted magnifications, time delays, and convergence/shear values for the four lensed SN images are reported in Table~\ref{tab:sie_tdelays_YS} (and also in Table~\ref{tab:lens_results}). We note that, strictly speaking, the BIC values can only be compared and combined when different models are constrained by the same data set, which does not apply to our results from three diferent bands. Nevertheless, the adopted weighting scheme is equivalent to weighting by the likelihoods, which is still a sensible treatment. 

\begin{table}[htbp]
\centering
\begin{tabular}{c|c|c|c|c|c}
\hline
 & & A & B & C & D \\
\hline
 & $\Delta x$ & $       0$ & $  0.1039 \pm   0.0049$ & $ -0.1059 \pm   0.0033$ & $ -0.2105 \pm   0.0047$ \\
 F475W & $\Delta y$ & $       0$ & $ -0.2351 \pm   0.0049$ & $ -0.2785 \pm   0.0031$ & $ -0.1123 \pm   0.0044$ \\
 & $m_{\rm AB}$ & $    22.9 \pm      0.1$ & $    23.8 \pm      0.1$ & $    22.9 \pm      0.1$ & $    23.7 \pm      0.1$ \\
\hline
 & $\Delta x$ & $       0$ & $  0.1011 \pm   0.0018$ & $ -0.1121 \pm   0.0012$ & $ -0.2187 \pm   0.0018$ \\
 F625W & $\Delta y$ & $       0$ & $ -0.2360 \pm   0.0017$ & $ -0.2817 \pm   0.0012$ & $ -0.1149 \pm   0.0017$ \\
 & $m_{\rm AB}$ & $    21.6 \pm      0.1$ & $    22.5 \pm      0.1$ & $    21.7 \pm      0.1$ & $    22.5 \pm      0.1$ \\
\hline
 & $\Delta x$ & $       0$ & $  0.1044 \pm   0.0011$ & $ -0.1110 \pm   0.0008$ & $ -0.2204 \pm   0.0012$ \\
 F814W & $\Delta y$ & $       0$ & $ -0.2345 \pm   0.0012$ & $ -0.2816 \pm   0.0009$ & $ -0.1137 \pm   0.0012$ \\
 & $m_{\rm AB}$ & $    20.5 \pm      0.1$ & $    21.5 \pm      0.1$ & $    20.7 \pm      0.1$ & $    21.4 \pm      0.1$ \\
\hline
\end{tabular}
\caption{Relative positions (with respect to Image A that has the highest S/Ns in all three bands and thus smallest positional uncertainties) and AB magnitudes measured by the {\tt lfit\_gui} approach in F475W, F625W, and F814W respectively. A constant 0.1 mag uncertainty is assumed. }
\label{tb:photometric_fit_YS}
\end{table}
\begin{table}
    \centering
    \begin{tabular}{c|c c c|c c c c}
    \hline
    Parameter & \multicolumn{3}{c|}{MAP} & \multicolumn{4}{c}{Marginalisation} \\
    \cline{2-4} \cline{5-8}
     & F475W & F625W & F814W & F475W & F625W & F814W & Combined \\
    \hline
    $\theta_{\rm E}$ [arcsec] & $0.1632$ & $0.1658$ & $0.1671$ & $0.164_{-0.002}^{+0.002}$ & $0.1659_{-0.0009}^{+0.0009}$ & $0.1671_{-0.0006}^{+0.0006}$ & $0.1664_{-0.0019}^{+0.0010}$ \\
    $q_{\rm m}$ & $0.647$ & $0.649$ & $0.624$ & $0.64_{-0.04}^{+0.05}$ & $0.648_{-0.017}^{+0.018}$ & $0.623_{-0.011}^{+0.012}$ & $0.632_{-0.019}^{+0.027}$ \\
    $\phi_{\rm m}$ [deg] & $69.4$ & $68.9$ & $69.1$ & $69.4_{-0.6}^{+0.6}$ & $68.9_{-0.2}^{+0.2}$ & $69.1_{-0.2}^{+0.2}$ & $69.0_{-0.3}^{+0.3}$ \\
   \hline
    \end{tabular}
    \caption{Key SIE model parameters from maximum a posteriori (MAP) estimation and marginalised posterior distributions in the {\tt lfit\_gui} approach.}
    \label{tab:sie_pars_YS}
\end{table}
\begin{table}
    \centering
    \begin{tabular}{c|cccc|cccc|c}
    \hline
    Image & \multicolumn{4}{c|}{Magnification} & \multicolumn{4}{c|}{Time delay [day]} & $\kappa$/$\gamma$ \\
    \cline{2-5} \cline{6-9} \cline{10-10}
     & F475W & F625W & F814W & Combined & F475W & F625W & F814W & Combined & Combined \\
    \hline
    A & $-2.2$ & $-2.2$ & $-2.0$ & $-2.0_{-0.3}^{+0.2}$ & $0.44$ & $0.45$ & $0.49$ & $0.48_{-0.05}^{+0.03}$ & $0.74_{-0.03}^{+0.02}$ \\
    B & $\phantom{-}4.1$ & $\phantom{-}4.2$ & $\phantom{-}3.9$ & $\phantom{-}4.0_{-0.2}^{+0.3}$ & $0.00$ & $0.00$ & $0.00$ & $0.00$ & $0.374_{-0.006}^{+0.009}$ \\
    C & $-3.7$ & $-3.7$ & $-3.3$ & $-3.5_{-0.4}^{+0.2}$ & $0.21$ & $0.21$ & $0.24$ & $0.23_{-0.02}^{+0.02}$ & $0.644_{-0.016}^{+0.011}$ \\
    D & $\phantom{-}4.6$ & $\phantom{-}4.6$ & $\phantom{-}4.2$ & $\phantom{-}4.4_{-0.2}^{+0.4}$ & $0.06$ & $0.06$ & $0.06$ & $0.06_{-0.01}^{+0.01}$ & $0.385_{-0.006}^{+0.009}$ \\
    \hline
    \end{tabular}
    \caption{Predicted magnifications, relative time delays, and $\kappa$/$\gamma$ values for the four lensed SN images by the {\tt lfit\_gui} approach (using the MAP estimation for individual bands and BIC-weighted average for the combined result). For the SIE model, $\kappa$ and $\gamma$ values are always the same.}
    \label{tab:sie_tdelays_YS}
\end{table}


\section{Modeling with \lenstronomy}
\lenstronomy~is a multi-purpose, open-source, community-lead, \textsc{Astropy}-affiliated gravitational lensing and image modeling package
\citep{birrer_lenstronomy_2018, birrer_lenstronomy_2021}\footnote{https://github.com/lenstronomy/lenstronomy}.
\lenstronomy~supports a large variety of lens models and surface brightness profiles, as well as multiple numerical options to treat point sources. The modularity of \lenstronomy~supports imaging modeling as well as catalogue-based model fitting. \lenstronomy~has been applied for time-delay cosmography of lensed quasars \citep{birrer_h0licow_2019,shajib_strides_2020} and lensed SNe as well as a variety of other lens modeling and image analysis applications \citep{gilman_warm_2020,shajib_dark_2021,schmidt_strides_2022}.

\subsection{The LS1 Method}

\begin{figure}[htbp]
\centering
\includegraphics[width=0.9\textwidth]{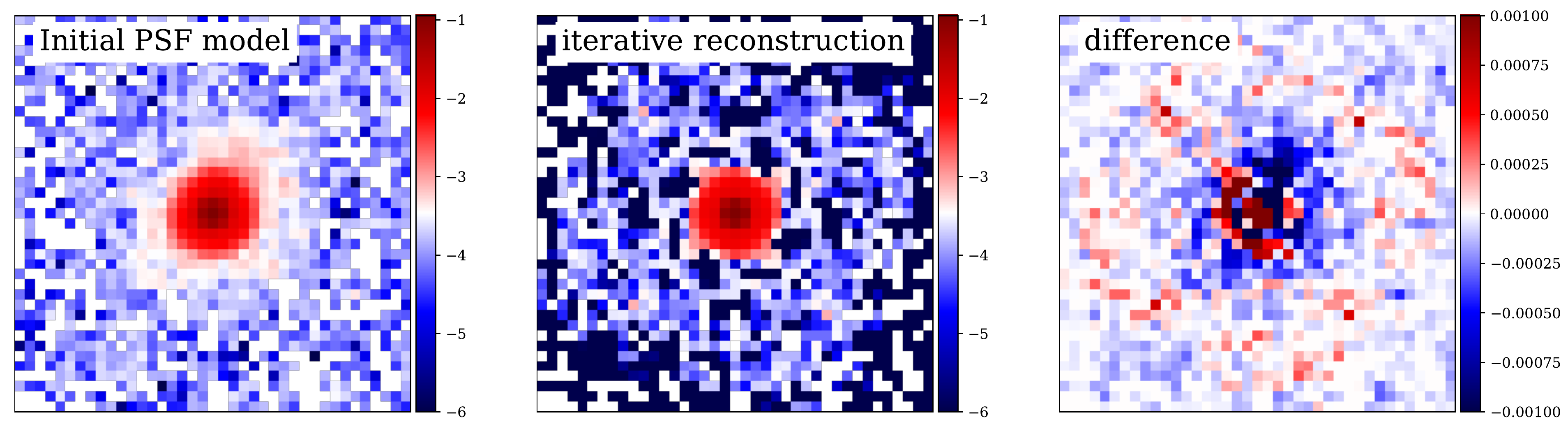}
\caption{The initial PSF model for the F814W band (left panel), the reconstruction after iterative improvement by \lenstronomy (middle panel), and the difference between the two (right panel).}
\label{fig:LS1_PSF}
\end{figure}

We described the mass profile of the lens galaxy by a singular isothermal ellipsoid \citep{kormann_isothermal_1994}, where the convergence is given by
\begin{equation}
\kappa(x, y) = \frac{\theta_\textrm{E}}{2 \sqrt{q_{\rm m}(x-x_{\rm m})^2+\frac{(y-y_{\rm m})^2}{q_{\rm m}}}}. 
\end{equation}
Here, $\theta_\textrm{E}$, $q_{\textrm{m}}$, and $(x_{\textrm{m}}, y_{\textrm{m}})$ are defined similarly as for equation~\ref{eq:kappa_spemd}. In order to fit the lens galaxy light profile, we stacked two \sersic profiles \citep{de_vaucouleurs_sur_1948, sersic_influence_1963}:
\begin{equation}
    I(R) = I_{\rm e}\exp{\left\{-b_n\left[\left(\frac{R}{R_{\rm e}}\right)^{1/n}-1\right]\right\}},
\end{equation}
where $I_{\rm e}$ is the intensity at the half-light radius $R_{\rm e}$. The constant $b_n$ is equal to $1.9992 n -0.3271$ \citep{birrer_lenstronomy_2018}, and $R\equiv\sqrt{q_{\textrm{S}} x^2 + y^2/q_{\textrm{S}}}$ with $q_{\textrm{S}}$ being the axis ratio of the \sersic profile. The supernova images were fitted as point sources on the image planes with a PSF model. We initiated the model fitting with the PSF model constructed by the \GLEE team and then further improved the PSF model using a built-in feature in \lenstronomy's that minimizes the residuals between the observed and reconstructed image around the supernova positions \citep{shajib_is_2019}. The comparison between the initial PSF model in the F814W band and the final reconstructed one is illustrated in Figure~\ref{fig:LS1_PSF}.
Additionally, we adopted a circular region around the lensing system for likelihood computation to avoid the boundary effect of the PSF convolution in the evaluated likelihood function.

\begin{figure}[htbp]
\centering
\includegraphics[width=0.92\textwidth]{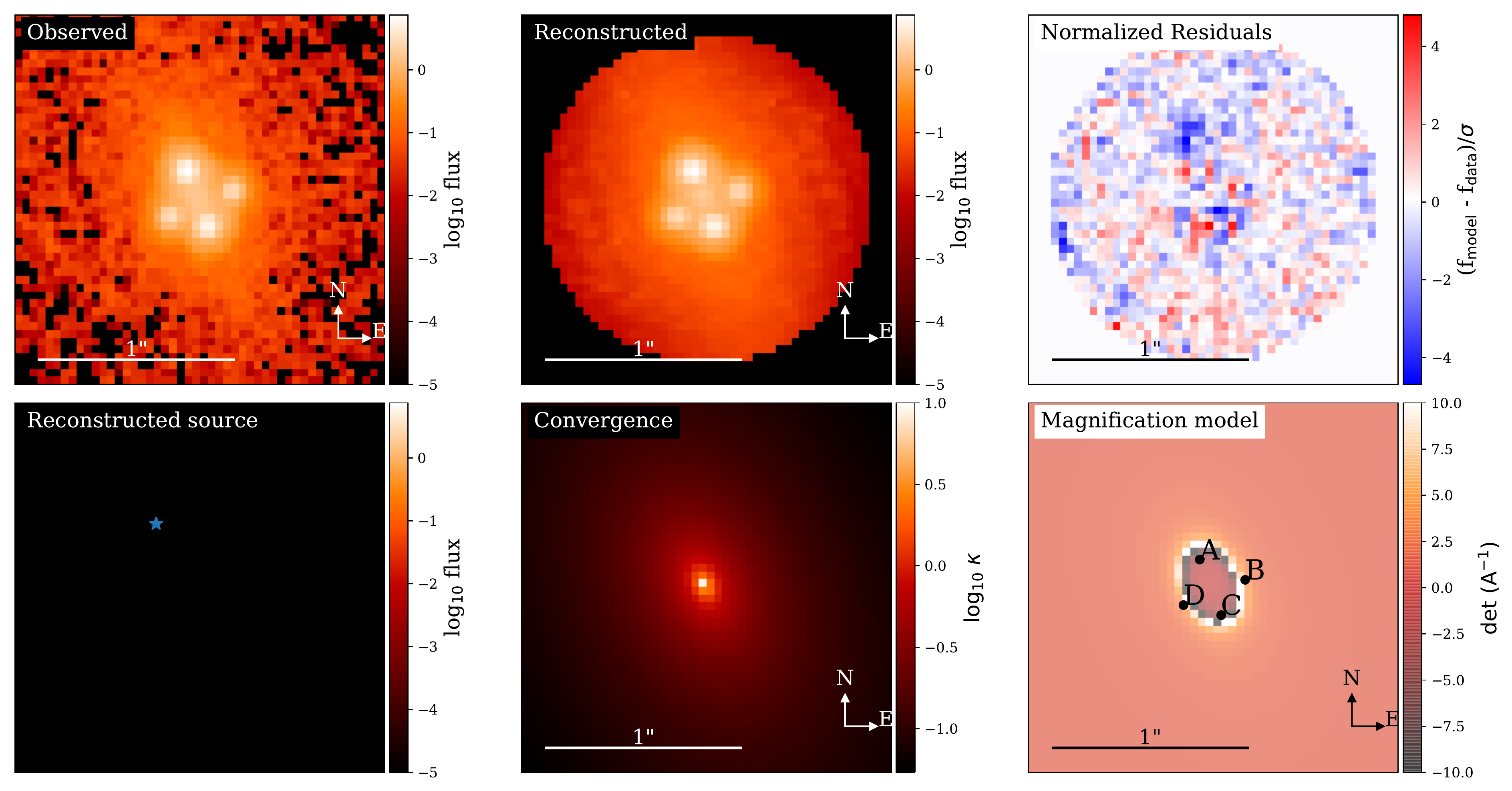}
\caption{The observed HST F814W-band data compared to the reconstructed model. \textit{Upper panel from left to right}: the observed image, the reconstructed light intensity, and the normalized residuals. The circular mask illustrates the image region used in computing the likelihood. \textit{Lower panel from left to right:} the reconstructed source, convergence (projected surface mass density), and magnification model with the four supernova images. In the reconstructed source panel, the blue star marks the unlensed position of the supernova, and no light from the host galaxy is detected above the noise level.}
\label{fig:LS1_reconstruction}
\end{figure}

We fitted the pixel-level data from the three optical HST bands in a joint likelihood. The uncertainties on the model parameters were obtained from a Markov Chain Monte Carlo (MCMC) sampling.
The flux ratios of the supernova images were not included in our lens model, because they failed to provide a good fit to the data and increased the reduced $\chi^2$ from $1.17$ to
$5.59$ (for the F814W filter). The reconstructed image model, source, convergence, and magnification model using the best-fit parameters from the converged MCMC chain are shown in Figure~\ref{fig:LS1_reconstruction}. Our measured $\kappa$ and $\gamma$ parameters at each image location are given in Table \ref{tab:ls1_kg}.

\begin{table}[h!]
    \centering
    \caption{Best-fit $\kappa/\gamma$ values for the LS1 lens model.}
    \label{tab:ls1_kg}
    \begin{tabular}{cc}
\toprule
Image&$\kappa/\gamma^a$\\
\hline
    A & $0.703^{+ 0.011}_{- 0.009}$\\
    B & $0.387^{+ 0.003}_{- 0.004}$\\
    C & $0.625^{+ 0.005}_{- 0.005}$\\
    D & $0.397^{+ 0.003}_{- 0.003}$\\
    \end{tabular}
    
    $^a$ Note that $\kappa=\gamma$ for an SIE model.
    
    \label{tab:my_label}
\end{table}

\clearpage

\subsection{The LS2 Method}
\begin{figure}[htbp]
\centering\includegraphics[width=0.92\textwidth, angle =270]{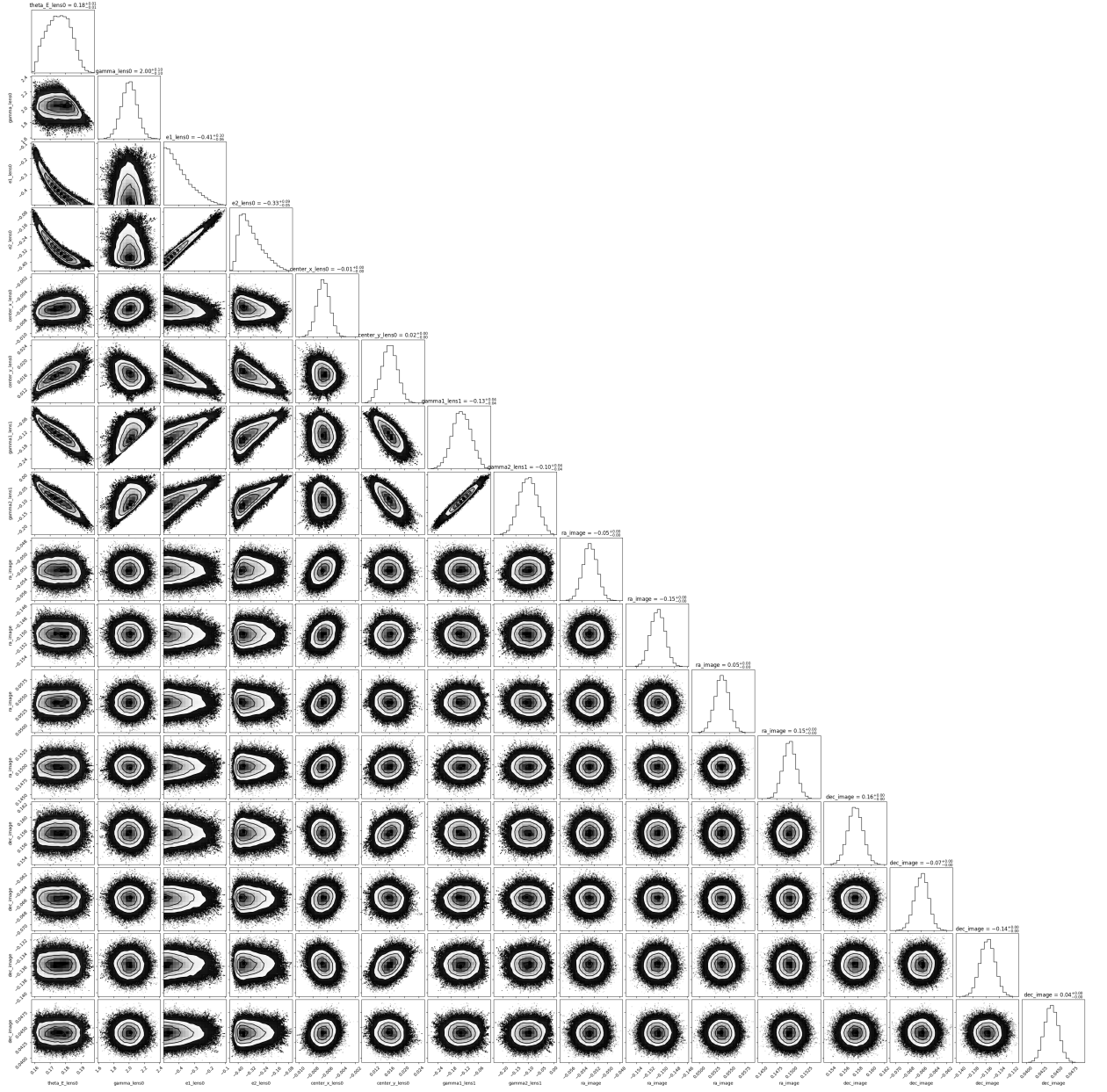}
\caption{Parameter sampling for the LS2 Lenstronomy method, for the F475W filter position measurements. }
\label{fig:LS2_parameters}
\end{figure}

\begin{figure}[htbp]
\centering\includegraphics[width=0.92\textwidth, angle =270]{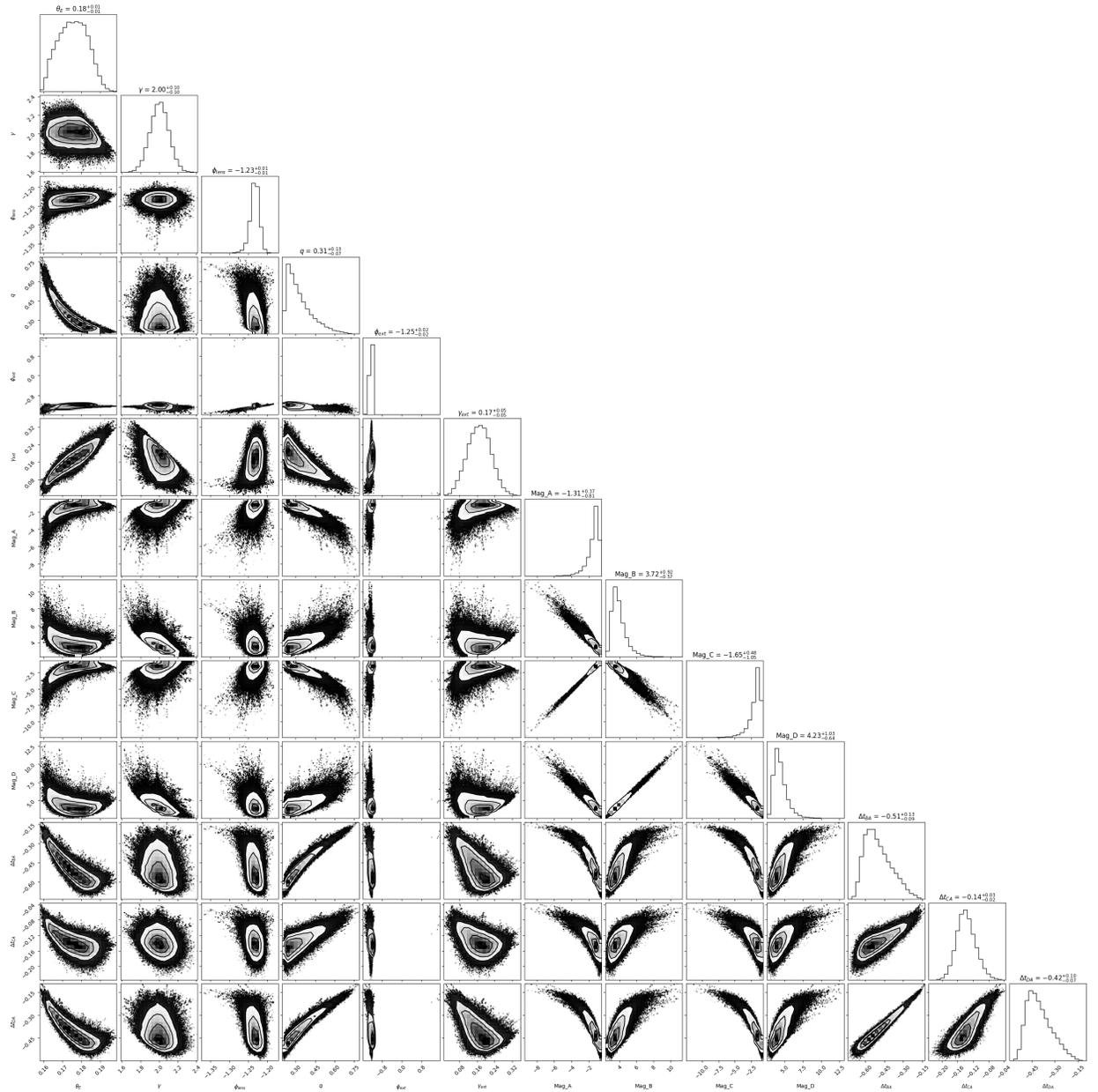}
\caption{The full corner diagram plot for the computed posterior estimates for the LS2 lens modeling fit, for the F475W filter position measurements.}
\label{fig:LS2_posteriors}
\end{figure}

The ``LS2'' team used the catalog-data modeling functionality of the
\lenstronomy software
package, using the
positions and positional uncertainties and redshifts reported in this paper. We adopted an elliptical power law 
mass profile \citep{tessore_elliptical_2015} plus external shear, and allowed all parameters to vary.  Models were
computed for each of the three WFC3/UVIS filters, F475W, F625W, and
F814W.  The MCMC parameter sampling for F475W is shown in
Figure~\ref{fig:LS2_parameters}, the posterior computed model parameters for the same filter are
shown in Figure~\ref{fig:LS2_posteriors}, and $\kappa,\gamma$ results are given in Table \ref{tab:ls2_kg}.  

\begin{table}[h!]
    \centering
    \caption{Best-fit $\kappa/\gamma$ values for the LS2 lens model.}
    \label{tab:ls2_kg}
    \begin{tabular}{ccc}
\toprule
Image&$\kappa$&$\gamma$\\
\hline
    A & $0.95^{+ 0.24}_{-0.20}$&$0.88^{+ 0.21}_{-0.16}$\\
    B & $0.31^{+0.07}_{-0.06}$&$0.43^{+ 0.04}_{-0.04}$\\
    C & $0.82^{+0.17}_{-0.15}$&$0.74^{+ 0.14}_{-0.11}$\\
    D & $0.32^{+0.07}_{-0.06}$&$0.43^{+ 0.04}_{-0.04}$\\
    \end{tabular}

    \label{tab:my_label}
\end{table}

\clearpage

\subsection{The SALT+LS Method}
\begin{figure}[htbp]
\centering\includegraphics[width=0.5\textwidth]{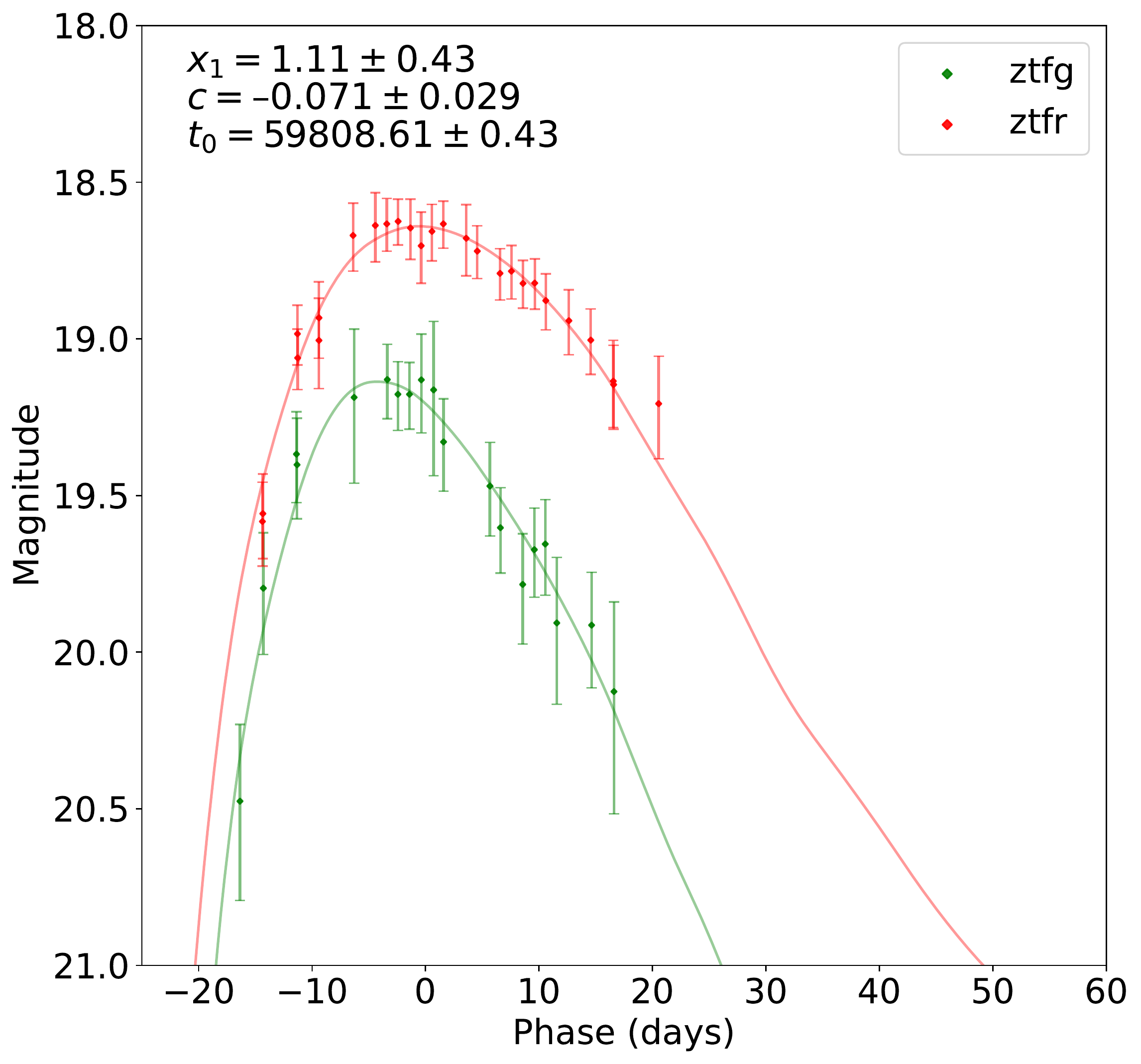}
\caption{ZTF $g$ and $r$ band light curves and best-fit SALT2.4 parameters.}
\label{fig:LC-SALT2}
\end{figure}

As a proof of concept, in lens modeling we use the expected SN Ia brightness as prior with a broad standard deviation of 0.3~mag.
First, using SNCosmo  \citep{barbary_sncosmo_2016}, we fit SALT2.4 for the publicly available ZTF photometric data in $g$ and $r$ bands.  
We obtain $x_1 = 1.11\pm0.43$, $c = -0.071 \pm 0.029$, with maximum light at MJD = $59808.54 \pm 0.43$ (Figure~\ref{fig:LC-SALT2}).  
These values are in good agreement with the best-fit SALT parameters from G22.
Note that G22 also used data from the Liverpool Telescope, which provided additional observations in $griz$ bands.  
To standardize the SN Ia brightness, we adopt peak $B$ magnitude of $M_B = -19.05$, $\alpha = -0.141$, and $\beta = 3.101$ \citep{betoule_improved_2014}.

\begin{figure}[htbp]
\centering\includegraphics[width=0.7\textwidth]{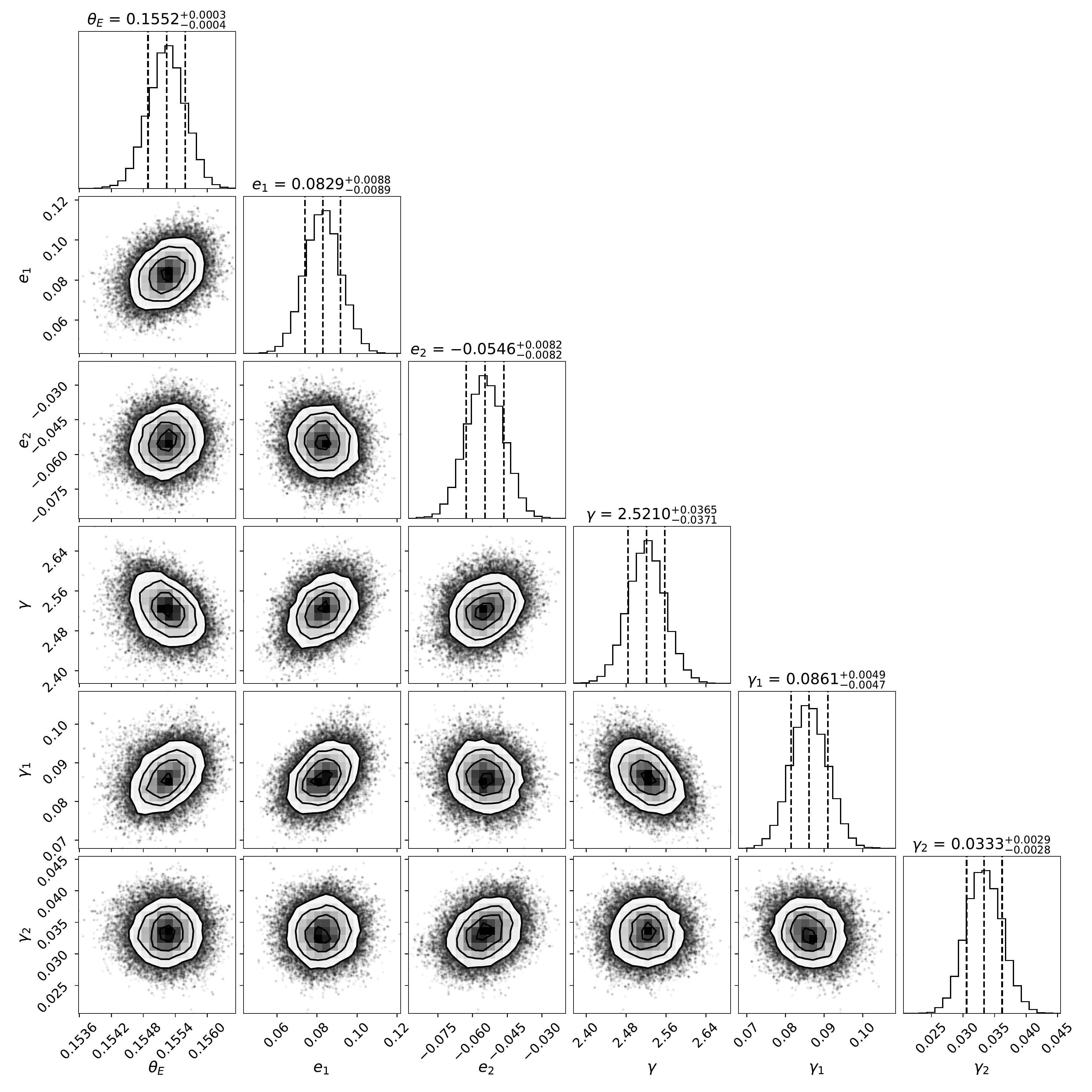}
\caption{A corner plot of the posterior samples for the parameters of the SALT+LS lens model.}
\label{fig:salt+ls-corner}
\end{figure}

\begin{figure}[htbp]
\centering\includegraphics[width=0.33\textwidth]{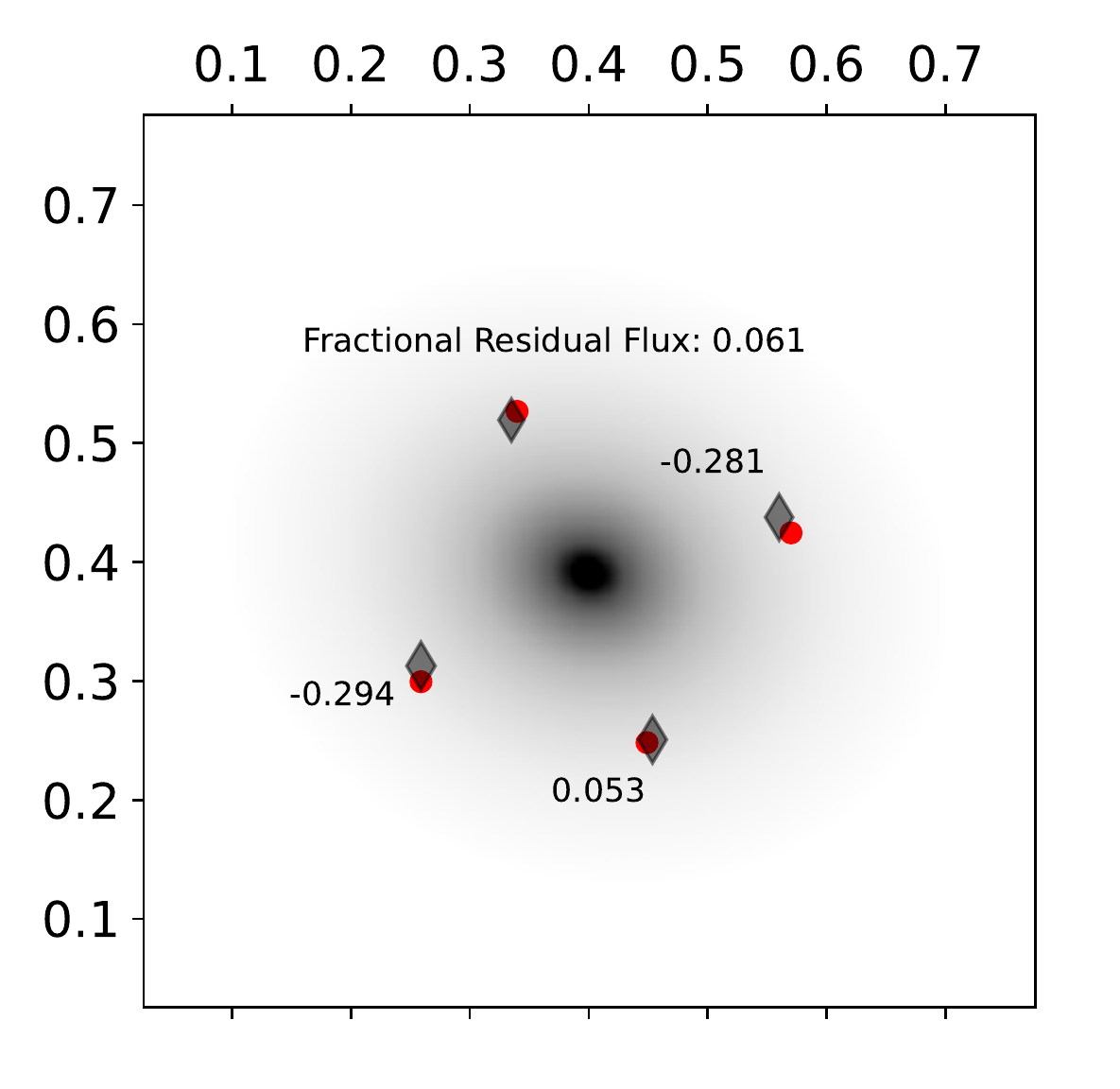}
\caption{The SALT+LS lens model prediction for image positions and fluxes. 
The red dots are the observed positions of the images and the gray diamonds are the model predicted positions.  
The number next to each image position is the fractional error of the model predicted flux.  
The tickmarks are in units of arcseconds.
The gray scale background shows the convergence ($\kappa$) map.
This image is shown in a slightly different orientation relative to most of the figures in this paper, 
but is the same as Figure~\ref{fig:LS1_reconstruction}. 
The brightest image, A, is still at the very top, and the other images are in the same alphabetical order clockwise from this reference point.}
\label{fig:lens_model_SALT2+LS}
\end{figure}

We combine the three HST optical bands by averaging the best-fit PSF centroids and adding the fluxes.
We assume a flat $\Lambda$CDM universe with $H_0=70$~km/s/Mpc and $\Omega_M=0.3$.  
Our model consists of an elliptical power law (EPL) main lens and external shear. 
Our loss function combines the summed squares of the difference in the delensed image positions and summed squares of the difference between the observed and the model predicted fluxes.  
The relative weight between the flux and position terms is adjusted to achieve the best overall fit. 
The results are summarized in Figure~\ref{fig:salt+ls-corner}.
We find a somewhat steep mass profile slope for the lensing galaxy: $\gamma_\mathrm{EPL} = 2.50$.
As G22 pointed out, with such a small $\theta_E$,
this system is in a regime of lensing galaxies that have seldom been studied before.  
Such systems can be used to probe the density profile at sub-kpc scales within the lensing galaxy core.
The image positions and model predictions agree to better than 
$0.023 \arcsec$ (Figure~\ref{fig:lens_model_SALT2+LS}).
The magnifications from this model for images B \& D (Table~\ref{tab:lens_results}) are in fairly good agreement with the corresponding expectations in Table~\ref{tab:td_mu}, whereas for images A \& C, the model predictions are lower by $< 2\,\sigma$ (see also Figure~\ref{fig:lens_model_SALT2+LS}). 
It appears that without taking into account microlensing and/or differential dust extinction, this is the best compromise the model can achieve. 
The total predicted magnification is 17.73.
Compared with the expectation of $24.3\pm2.7$ from G22, 
it is smaller by $2.5\,\sigma$.

We find that without using SN~Ia brightness as prior, it is possible to find models with acceptable predictions for both image positions and flux values, 
but they tend to have a much shallower slope ($\gamma_\mathrm{EPL} \lesssim 1.75$).  
In contrast, using the SN Ia brightness prior, we find $\gamma_\mathrm{EPL}$ to be consistently $\gtrsim 2.5$, 
whether we use single band data or combine the different bands.
We also note that if the host galaxy identification in G22 is correct, 
this system is possibly in a unique situation in that the core of the host galaxy is not multiply-imaged. 
This makes the modeling of this system especially challenging: we cannot separately perform lens modeling using the lensed host galaxy in contrast to the other small-$\theta_E$ lensed SN~Ia \citep{dhawan_magnification_2019}.

We now briefly compare the SALT+LS model with the other four models presented in this paper that only use image positions. 
With regard to $\theta_E$, when fluxes are taken into consideration by the \GLEE team, they have found acceptable models with $\theta_E$ in agreement with the SALT+LS best-fit value.
We further note that the time delay predictions from the SALT+LS model are in good agreement with those from the \GLEE model that has a similar $\theta_E$.
With regard to magnification: 1) The total magnification from the these four models ranges from 9.2 to 15.7 (Table~\ref{tab:lens_results}).  
The SALT+LS model predicts a magnification of 17.43, higher than these four models.
2) Whereas the SALT+LS model predicts the magnifications for the brighter two images, A and C, to be higher than those of B and D, 
the other four models predict the opposite.
And yet, as mentioned before, even the total magnification from the SALT+LS model is $\sim 30\%$ 
lower than the expected magnification from G22 or from Table~\ref{tab:td_mu} based on SN~Ia brightness.
The ($\kappa$, $\gamma$) values at the locations of the images are, in order of A-D: 
(0.25, 0.86), (0.22, 0.56), (0.26, 0.88), and (0.21, 0.60).  
Given that this appears to be a fairly normal SN~Ia, it is possible that microlensing and/or differential dust extinction (likely to be small, given the small color differences for the four images shown in Table~\ref{tab:im_mags}) have played a significant role.
If so, for the SALT+LS model, the optimization can be distorted in a way to compensate for these effects.
Thus, for the uncertainties for $\kappa$ and $\gamma$, we report the largest uncertainties for the four images, which are 0.024 and 0.023 respectively (Table \ref{tab:salt_ls_kg}).
Once follow-up HST observations are completed after the SN has faded and improved photometry has been obtained, we will revisit this model.

\begin{table}[h!]
    \centering
    \caption{Best-fit $\kappa/\gamma$ values for the SALT+LS lens model.}
    \label{tab:salt_ls_kg}
    \begin{tabular}{ccc}
\toprule
Image&$\kappa$&$\gamma$\\
\hline
    A & $0.25\pm0.024$& $0.86\pm0.023$\\
    B & $0.22\pm0.024$& $0.56\pm0.023$\\
    C & $0.26\pm0.024$& $0.88\pm0.023$\\
    D & $0.21\pm0.024$& $0.60\pm0.023$\\
    \end{tabular}

    \label{tab:my_label}
\end{table}

\end{document}